\documentclass{ethpaper}
\usepackage{graphicx}
\usepackage{here}
\usepackage{amssymb}
\usepackage{amsmath}
\usepackage{subfigure}
\usepackage{lineno}

\begin{document}
\begin{titlepage}
\ethnote{}
\title{Proof-of-principle of a new geometry 
\\for sampling calorimetry\\
using inorganic scintillator plates
}
\begin{Authlist}
R. Becker, G. Dissertori, A. Gendotti, Q. Huang, D. Luckey\footnote{also at Massachusetts Institute of Technology, Cambridge, Massachusetts, USA},
W.~Lustermann, S. Lutterer,\\\underline{F. Nessi-Tedaldi}, F. Pandolfi, F. Pauss, M. Peruzzi, M. Quittnat, R.~Wallny
\Instfoot{eth}{Institute for Particle Physics, ETH Zurich, 8093 Zurich, Switzerland}
\end{Authlist}
\maketitle
\begin{abstract}
A novel geometry for a sampling calorimeter employing inorganic scintillators as an active medium is presented. To overcome the mechanical challenges of construction, an innovative light collection geometry has been pioneered, that minimises the complexity of construction. First test results are presented, demonstrating a successful signal extraction.
The geometry consists of a sampling calorimeter with passive absorber layers interleaved with layers of an active medium made of inorganic scintillating crystals. Wavelength-shifting (WLS) fibres run along the four long, chamfered edges of the stack, transporting the light to photodetectors at the rear. To maximise the amount of scintillation light reaching the WLS fibres, the scintillator chamfers are depolished.  It is shown herein that this concept is working for cerium fluoride (CeF$_3$) as a scintillator. Coupled to it, several different types of materials have been tested as WLS medium. In particular, materials that might be sufficiently resistant to the High-Luminosity Large Hadron Collider radiation environment, such as cerium-doped Lutetium-Yttrium Orthosilicate (LYSO) and cerium-doped quartz, are compared to conventional plastic WLS fibres. Finally, an outlook is presented on the possible optimisation of the different components, and the construction and commissioning of a full calorimeter cell prototype is presented.\end{abstract}
\vspace{5.5cm}
\conference{\em To be published in Proceedings CALOR 2014,\\the 16th International Conference on Calorimetry for High-Energy Physics (Gie{\ss}en, Germany)}
\end{titlepage}
\section{Introduction}
The detectors now in operation at the Large Hadron Collider (LHC) at CERN  will have to
face a challenging environment after the accelerator upgrade to
High-Luminosity running (HL-LHC), that is currently being planned to start in 2022.
The radiation levels expected there are typically a factor 10 larger, and the integrated luminosities ---  thus integrated radiation levels --- a factor 6 more important than the design values at the LHC. The HL-LHC design parameters at 14 TeV foresee a luminosity of $5\times 10^{34}\;\mathrm{cm^{-2}\;s^{-1}}$, up to a total integrated luminosity of
$30\times 10^5\;\mathrm{pb^{-1}}$~\cite{r-LUC}. Typical exposure levels in the high-rapidity region of the calorimeters reach 30 Gy/h in ionising radiation and
$2\times 10^{14}\;\mathrm{cm^{-2}}$ in integrated fluences of high-energy hadrons. Calorimeters designed for an adequate performance through the years of running at the LHC may need to be partially upgraded, to withstand these increased exposure levels at the HL-LHC.
\section{A new sampling calorimeter geometry}
\label{s-GEO}
Sampling calorimeters are being considered with a renewed interest as an option for HL-LHC detector upgrades.
\begin{figure}[!t]
\centering
\includegraphics[width=12cm]{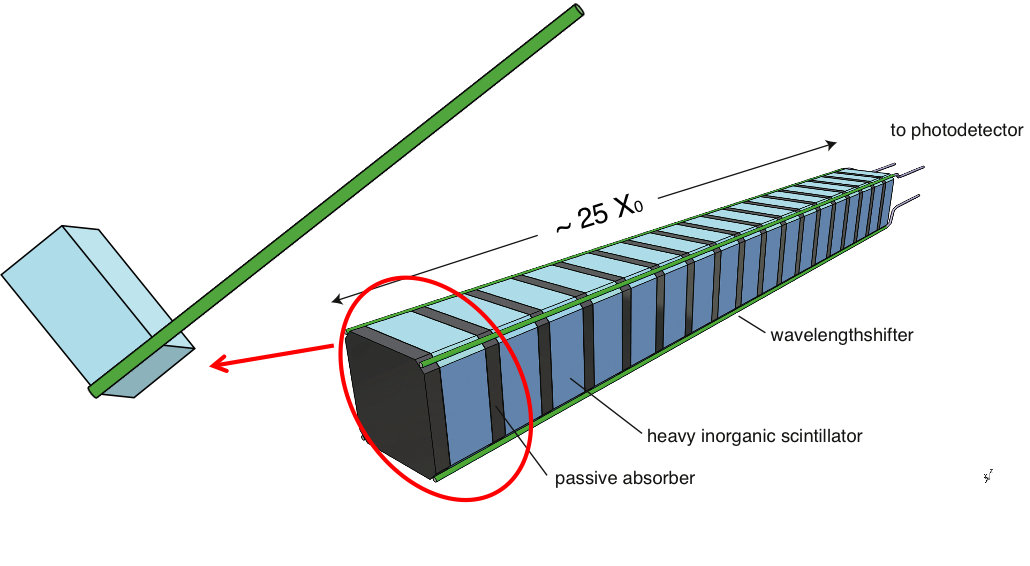}
\caption{Concept drawing for an innovative sampling calorimeter cell geometry using inorganic scintillator plates as an active material, with wavelength-shifting fibres running along the four long cell edges, and enlarged view on the components used for proof-of-principle measurements.}
\label{f-channelelement}
\end{figure}
Some inorganic scintillators have been identified, that are likely to perform adequately in the strong radiation field and high particle fluences present during HL-LHC running (\cite{r-NIMCEF3},\cite{r-ZHULYSO}). A sampling calorimeter employing them might offer an interesting option in terms of costs and short light path. However, inorganic crystalline materials present mechanical challenges compared to plastic scintillators: this paper describes an innovative light collection geometry that minimises the mechanical processing complexity. It presents first test results demonstrating a successful signal extraction and evaluates possible WLS candidates.

\begin{figure}[!b]
\includegraphics[width=10pc]{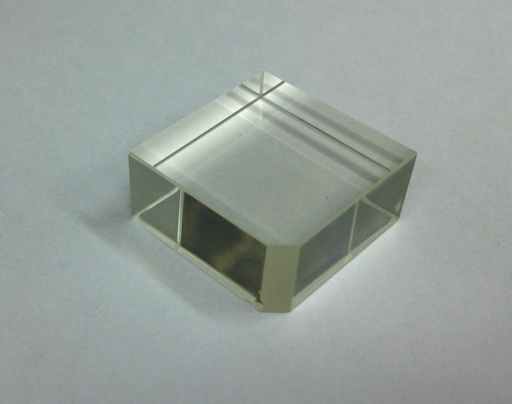}\hspace{2pc}%
\begin{minipage}[b]{22pc}\caption{\label{f-cef3}The CeF$_3$ crystal sample ($20\times 20\times 10\;\mathrm{mm}^3$) used in this test, provided with a 3 mm wide, depolished chamfer along one of its 10 mm long edges.}
\end{minipage}
\end{figure}

The geometry (Fig.~\ref{f-channelelement}) consists of a sampling calorimeter cell made of passive absorber layers (e.g.~W or Pb) interleaved with layers of an active medium made of inorganic scintillating crystals. Wavelength-shifting (WLS) fibres run along the four long edges of the stack, transporting the light to photodetectors at the rear. The chamfers along those edges have a width matching the WLS fibre dimensions. On the crystal samples they are depolished to maximise the amount of scintillation light escaping the scintillator and reaching the fibres. The simplicity of the design minimises the mechanical processing complexity, compared to other geometries with grooves or holes machined through the stack. 

In this paper, a study has been performed, in order to to show that this concept is working for cerium fluoride (CeF$_3$)~\cite{r-NIMCEF3} used as the scintillating material.
For the test, a CeF$_3$ crystal has been prepared, with a 3 mm wide and depolished chamfer, as visible in Fig.~\ref{f-cef3}. Coupled to it, several different types of materials have been tested as WLS medium, using cosmic muons and laboratory $\gamma$ sources to excite the CeF$_3$ scintillation. In particular, materials that might be sufficiently resistant to the HL-LHC radiation environment, such as cerium-doped Lutetium-Yttrium Orthosilicate (LYSO)  and cerium-doped quartz, are compared to conventional plastic WLS fibres. The most promising material being studied as a  WLS candidate is in fact cerium-doped quartz, which is photo-luminescent in range of wavelengths of the CeF$_3$ scintillation emission~\cite{r-vedda}.

The presented results provide a proof-of-principle that this kind of light extraction geometry via WLS fibres is viable. Finally, an outlook is presented on a further possible optimisation of the different components and the steps leading to a full calorimeter cell prototype construction are shown.

\section{Cerium fluoride as a scintillator for HL-LHC calorimetry}
\label{s-CEF}
Cerium fluoride is a scintillating crystal whose density ($\rho=6.16\; {\mathrm{g/cm}}^3$), radiation length ($X_0 =
1.68$ cm), Moli\`ere radius ($R_M = 2.6$ cm), nuclear interaction length
($\lambda_I = 25.9$ cm) and refractive index ($n = 1.68$) make it a
competitive medium for compact calorimeters. For Barium-doped crystals, the emission is centred
at 340 nm, with decay time constants of 10 - 30~ns; it is insensitive
to temperature changes (${\mathrm{dLY/dT}}\; (20^o$C$)=0.08\%/^o$C) as well as bright (4 -
10\% of NaI($T\ell$)) and thus suitable for high-rate, high-precision
calorimetry~\cite{r-AND2,r-EACEF3}. Minimum ionising particles deposit $7.9$ MeV/cm in CeF$_3$, and cosmic $\mu$ are thus an interesting means of exciting its scintillation. It should be noted that cerium is a readily available
rare earth, which would allow moderating raw material costs, were a mass production envisaged. Its melting point of $1460\;^o$C allows
applying well-known crystal growth technologies. 
A feature crucial for use in calorimetry at the HL-LHC, is its performance under exposure to considerable fluxes of high-energy hadrons. It has been demonstrated, that crystals can be grown, where the damage fully recovers~\cite{r-NIMCEF3}, and thus never builds up as it is the case for lead tungstate~\cite{r-LTNIM} and for LYSO~\cite{r-NIMLYSO}.
\section{The test setup layout and its commissioning}
\label{s-STP}
\begin{figure}[!b]
\begin{center}
 \begin{tabular}[h]{ccc}
{\mbox{\includegraphics[width=42mm]{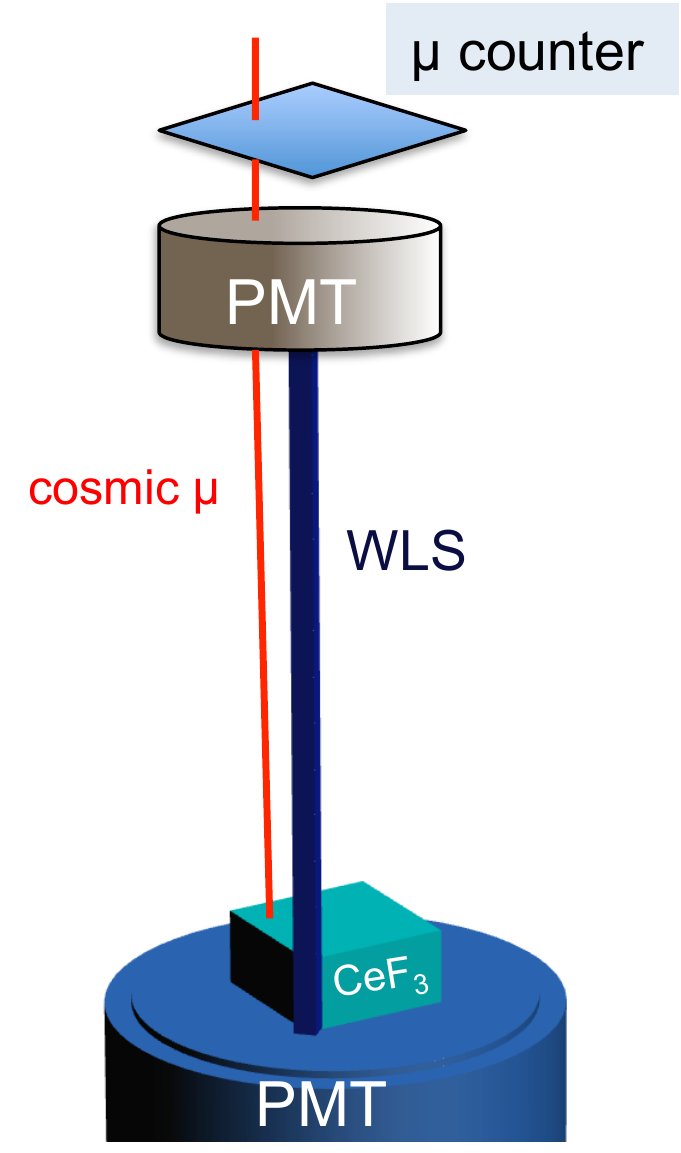}}} &
{\mbox{\includegraphics[width=40mm]{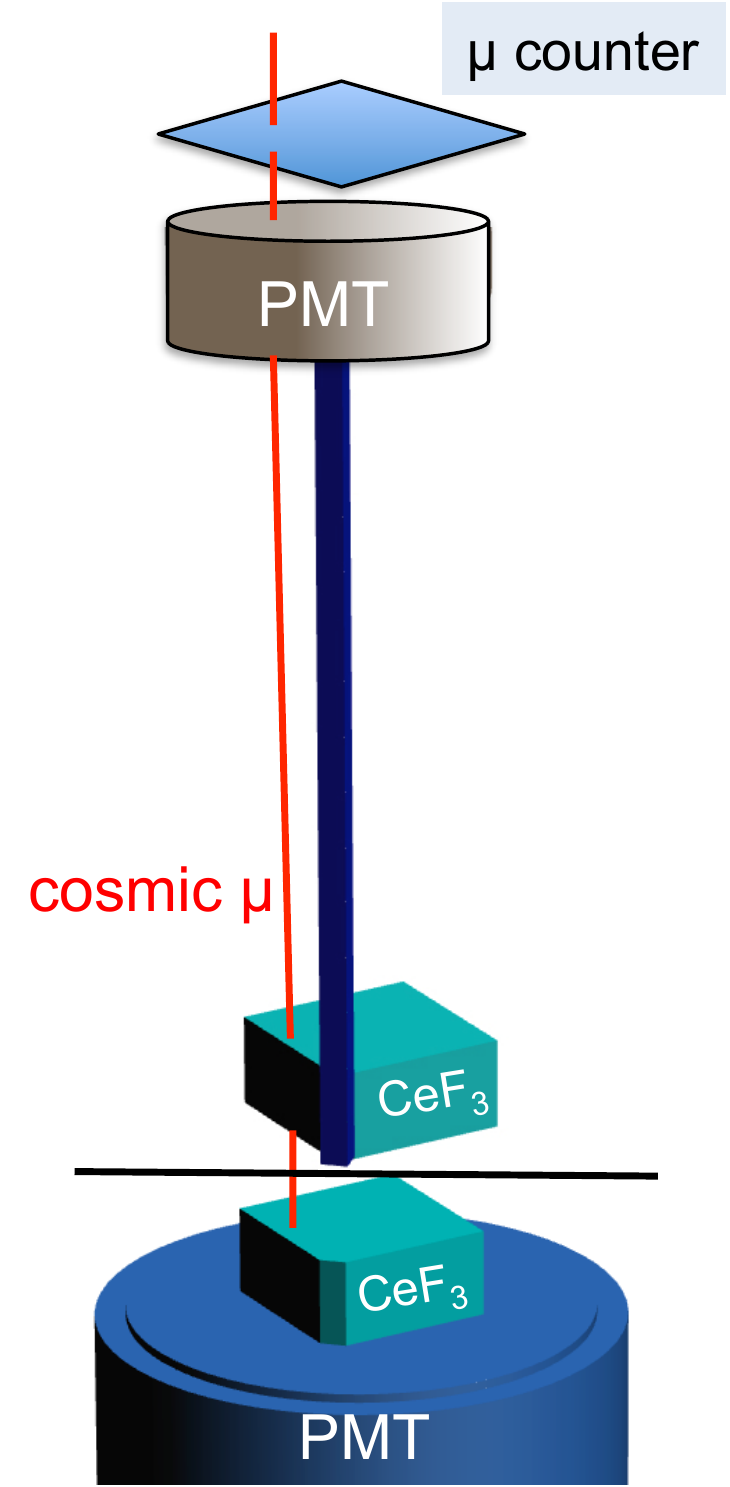}}} &
{\mbox{\includegraphics[width=40mm]{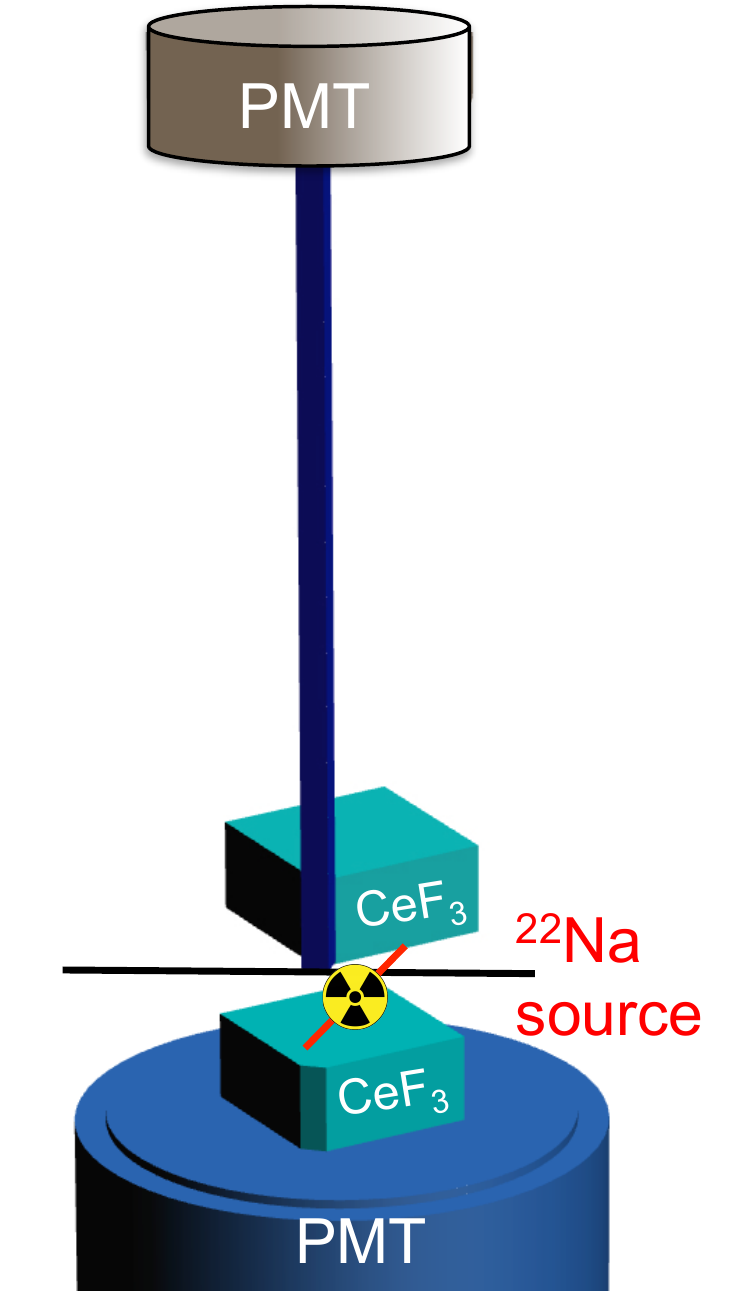}}}
   \end{tabular}
\end{center}
\caption{Test setup configurations: Left: setup 1, where a CeF$_3$ crystal is both, coupled to the bottom PMT for a direct scintillation light output measurement, and to a WLS bar seen by the top PMT. Center: setup 2,  where the bottom CeF$_3$ crystal is coupled to a PMT, and the top crystal is coupled to a WLS bar. Right: setup 3 is used to trigger on annihilation photons from a $^{22}$Na source placed between the crystals. Optical screens are used to optically insulate the crystals from one another and from the top PMT.}
\label{f-setup}
\end{figure}
\begin{figure}[!t]
\includegraphics[width=12pc]{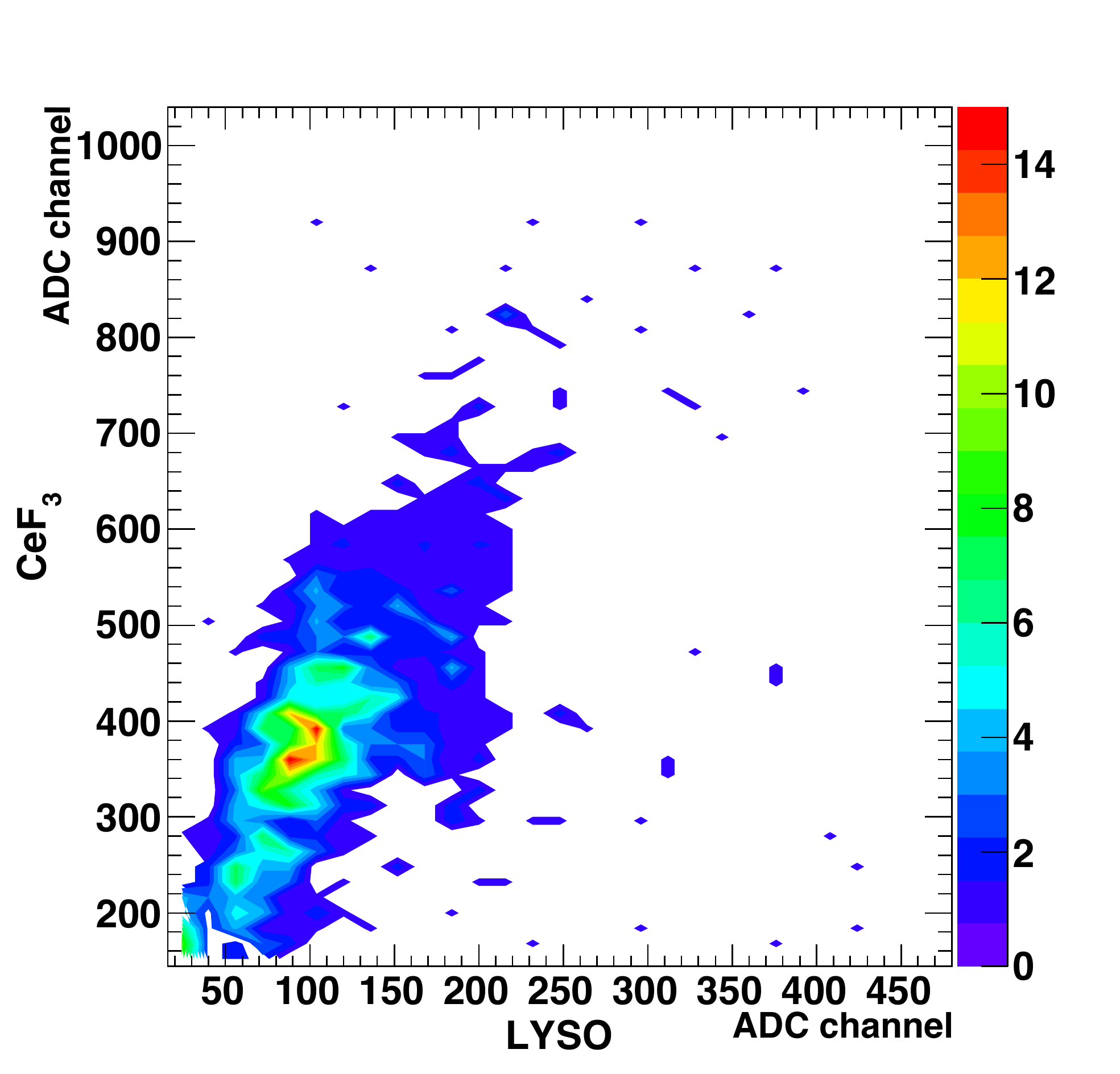}\hspace{2pc}%
\begin{minipage}[b]{24pc}\caption{\label{f-mu1s}For cosmic muons in setup 1 (Fig.~\ref{f-setup}), correlation between direct scintillation signal from CeF$_3$ and WLS signal from the LYSO bar.}
\end{minipage}
\end{figure}

For the presented proof-of-principle studies, a few variations of the same laboratory setup have been used, as visible in Fig.~\ref{f-setup}.
The first setup (Fig.~\ref{f-setup}, left) foresees one of the big crystal faces  to be coupled to a photomultiplier (PMT) of type Hamamatsu H7415 provided with a bialkali photocathode. On all other faces, the crystal was wrapped in Teflon, while the 3 mm chamfer was coupled, via an air gap, to a LYSO bar used here as a wavelength-shifter (WLS), due to its photoluminescence characteristics. The LYSO bar was parallelepipedic, with dimensions $3\times 3\times 100\;\mathrm{mm}^3$
yielding a volume of $0.9\;\mathrm{cm}^3$, of the same order of magnitude as the crystal volume of $4\;\mathrm{cm}^3$. At one end, the naked LYSO bar
was coupled to a bialkali-photocathode PMT of type Hamamatsu H8443, while the opposite end was aluminised. Optical screens were introduced, to make sure that no direct scintillation light from the CeF$_3$ could reach the PMT coupled to the LYSO bar.
The PMT output from CeF$_3$ scintillation light was split into two. One signal was used to trigger for cosmic muons traversing the crystal, in coincidence with the
signal from a plastic scintillator placed on top of the setup, with $40\times 40\;\mathrm{mm}^2$ active area. The other signal was integrated with an Analogue-to-Digital converter (ADC).
The PMT signal from the LSYO light output was integrated by a different ADC, and data were acquired on an event-by-event basis.
A correlation was observed from this setup, for a measurement with cosmic muons, which is visible in Fig.~\ref{f-mu1s}. It is apparent that a WLS signal due to the LYSO photoluminescence is detected. 
\begin{figure}[!b]
\begin{center}
 \begin{tabular}[h]{ccc}
{\mbox{\includegraphics[width=52mm]{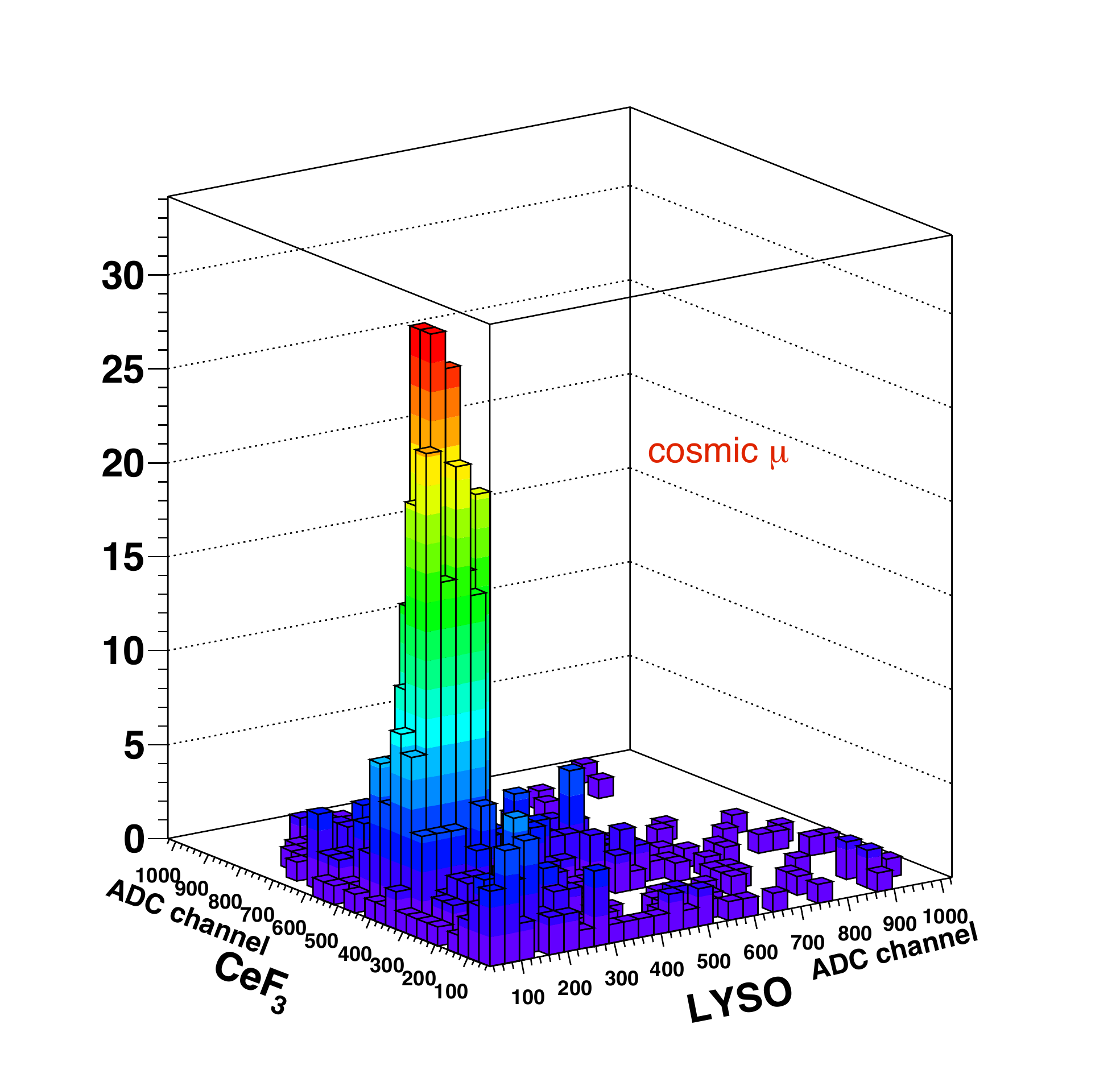}}} &
{\mbox{\includegraphics[width=50mm]{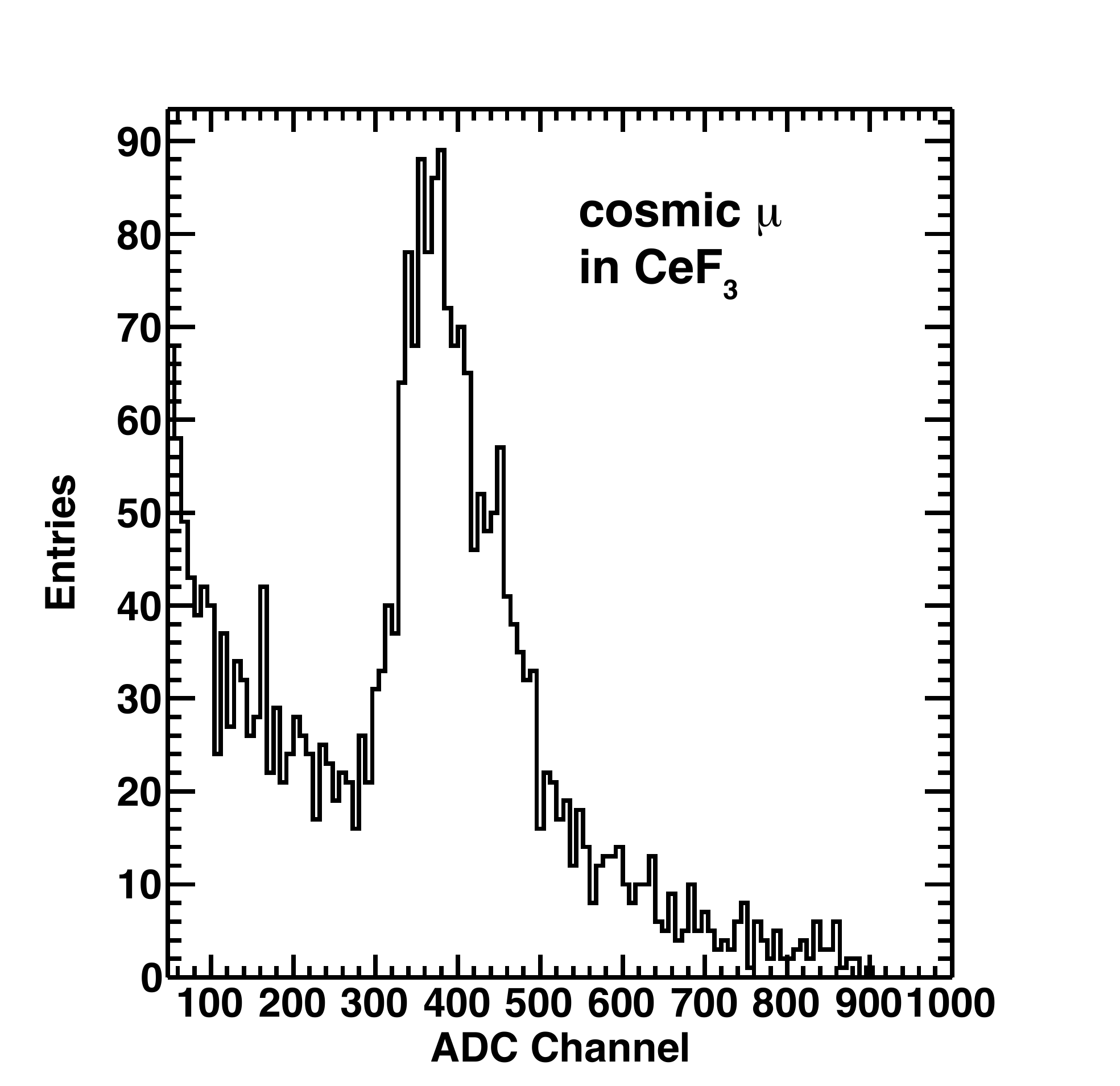}}} &
{\mbox{\includegraphics[width=50mm]{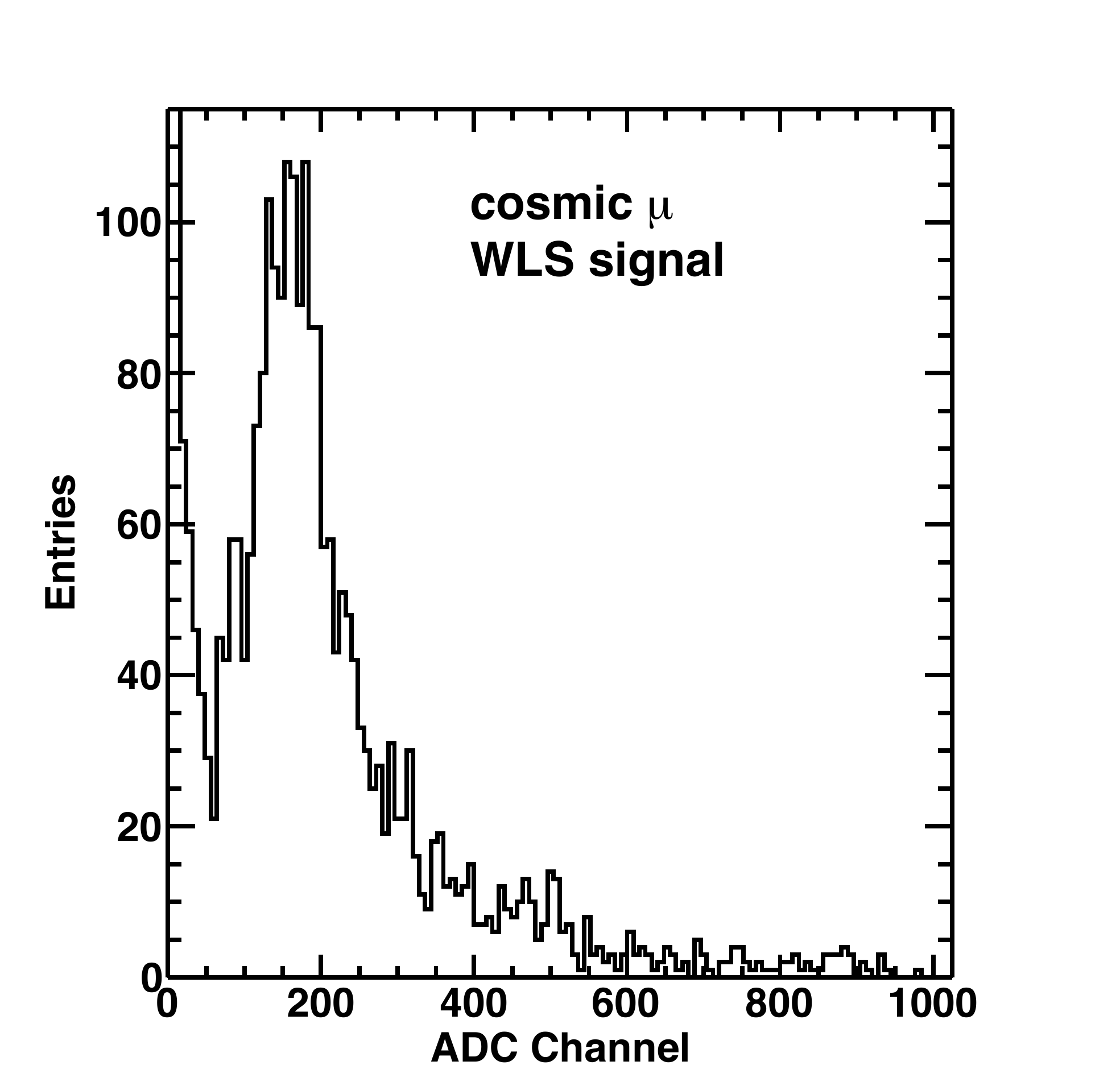}}}
   \end{tabular}
\end{center}
\caption{For cosmic muons  in setup 2 (Fig.~\ref{f-setup}), correlation between direct scintillation signal from the bottom CeF$_3$ and WLS signal from the LYSO bar coupled to the top CeF$_3$ crystal, and one-dimensional projections of the distributions.}
\label{f-setup2}
\end{figure}
\begin{figure}[!t]
\begin{center}
 \begin{tabular}[h]{cc}
{\mbox{\includegraphics[width=50mm]{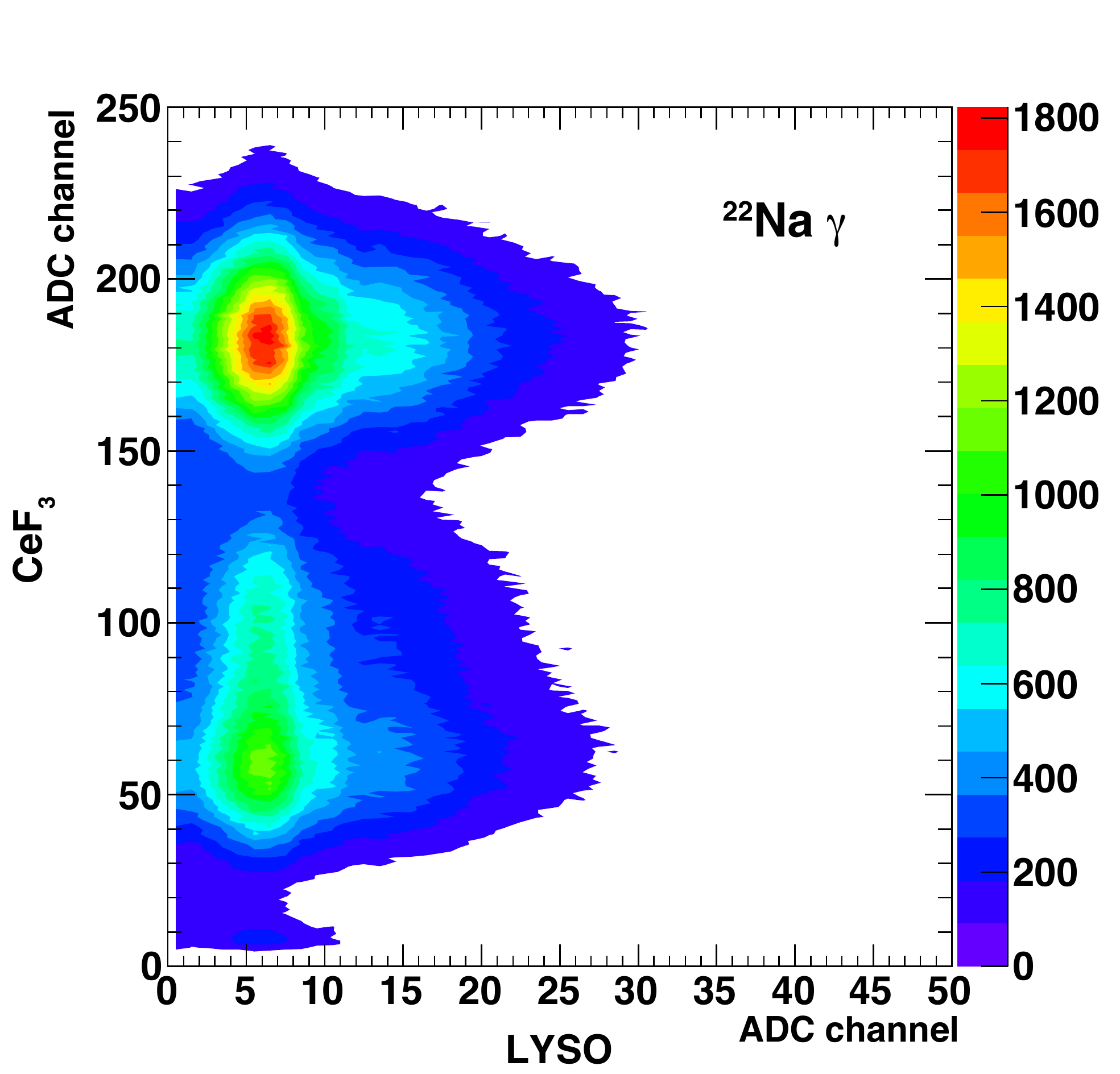}}} &
{\mbox{\includegraphics[clip=true, trim= 0cm 8mm 5mm 0cm,width=100mm]{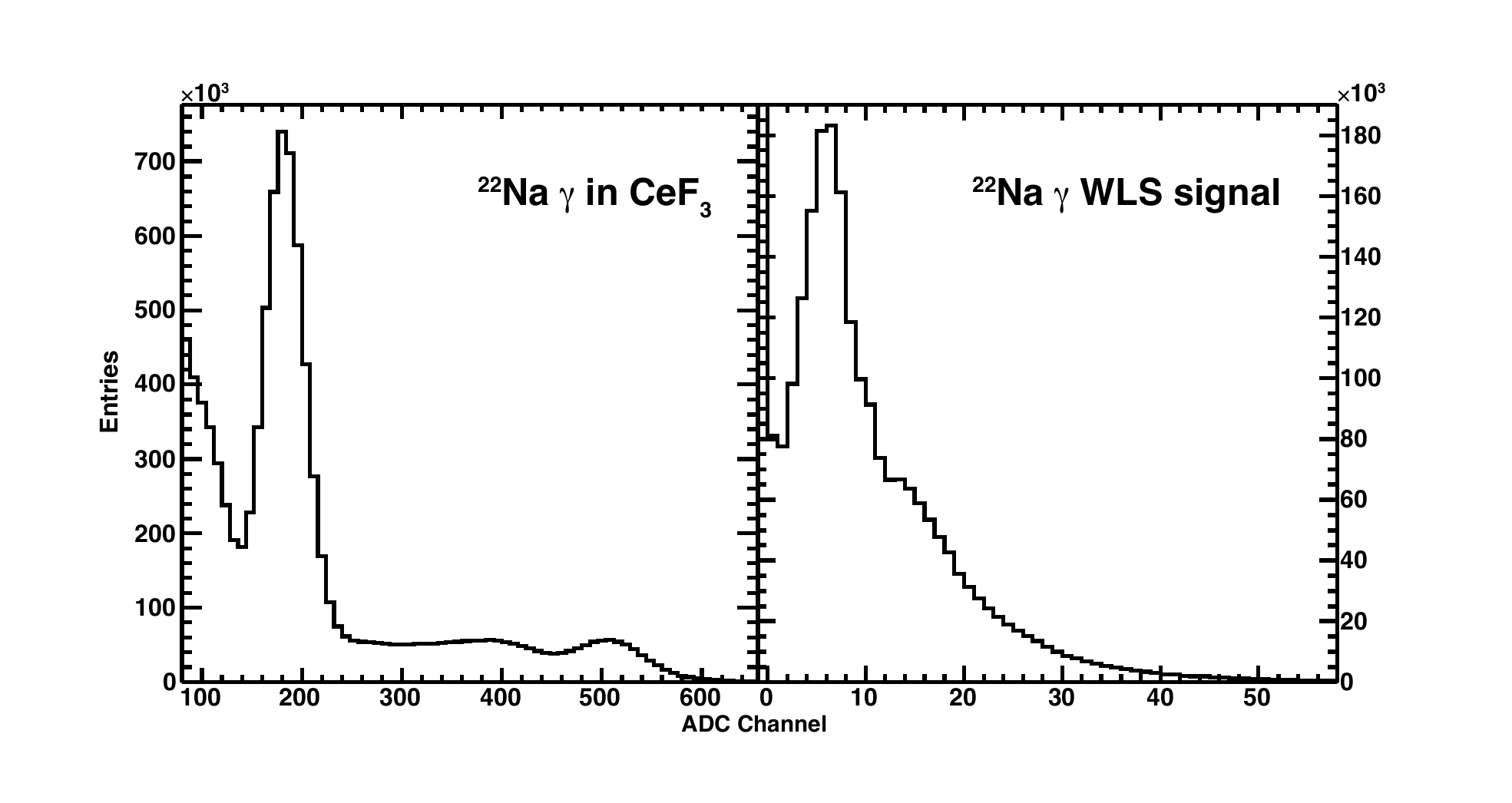}}}
   \end{tabular}
\end{center}
\caption{For a $^{22}$Na source in setup 3  (Fig.~\ref{f-setup}), correlation between direct scintillation signal from the bottom CeF$_3$ and WLS signal from the LYSO bar coupled to the top CeF$_3$ crystal, and one-dimensional projections of the distributions.}
\label{f-setup3}
\end{figure}

However, the setup 1 suffers from the drawback that part of the scintillation light from the CeF$_3$ crystal is not available for producing a WLS signal, because it is directly detected.
To overcome this limitation, setup 2 was adopted, with an identical, second CeF$_3$ crystal for direct scintillation light detection, to make all the light from the top crystal available for wavelength-shifting. Both crystals were wrapped in teflon, except for the side coupled to the PMT or the WLS bar, and they were optically isolated from each other. The resulting correlation and spectra when triggering for cosmic muons are shown in Fig.~\ref{f-setup2}. 
With a gain $G=1.5\times10^7$, a PMT quantum efficiency of 25\% in average over the CeF$_3$ emission spectrum, and an attenuation of 35~db on the signal, a light output of 300~p.e./MeV has been obtained, corresponding to a light yield $S_{CeF_3}=1200~\gamma$/MeV. For the PMT coupled to the LYSO bar, the gain used was $1.3\times10^7$ and the average quantum efficiency 20\%, yielding a WLS light output for this setup of  $S_{WLS}=2.4$~p.e./MeV, thus a light yield of 12~$\gamma$/MeV and a WLS to direct light yield ratio $R_{\gamma}=1\%$.

A study has also been performed placing a $^{22}$Na source between the CeF$_3$ crystal, triggering on back-to-back 511 keV annihilation $\gamma$ quanta (setup 3 in Fig.~\ref{f-setup}). The correlation and spectra are found in Fig.~\ref{f-setup3}. In the CeF$_3$ spectrum, one distinguishes well the photoelectric peak from 511 keV annihilation photons, and the Compton background and photoelectric peak for the 1.275 MeV emission which is occasionally detected in coincidence with a back-to-back annihilation photon. The distribution for WLS signals is less striking, due to the small amplitudes involved, but the photoelectric peak and the background at higher ADC channels are clearly distinguishable. One infers from this measurement a  light output of 230~p.e./MeV and a light yield $S_{CeF_3}=940~\gamma$/MeV. For the PMT coupled to the LYSO bar,  one obtains a WLS light output for this setup of  $S_{WLS}=1.4$~p.e./MeV, thus a light yield of 7~$\gamma$/MeV and a WLS-to-direct light yield ratio $R_{\gamma}=0.8\%$. Taking into account possible systematic effects due to differences in light collection between the two cases, one can consider the results for cosmic muons and $\gamma$ quanta reasonably compatible.

\begin{figure}[!h]
\begin{minipage}[b]{24pc}\caption{\label{f-LYSOscint}For a $^{22}$Na source in setup 3  (Fig.~\ref{f-setup}), correlation between direct scintillation signal from the bottom CeF$_3$ and direct scintillation signal from the LYSO bar (see text for details).}
\end{minipage}
\includegraphics[width=12pc]{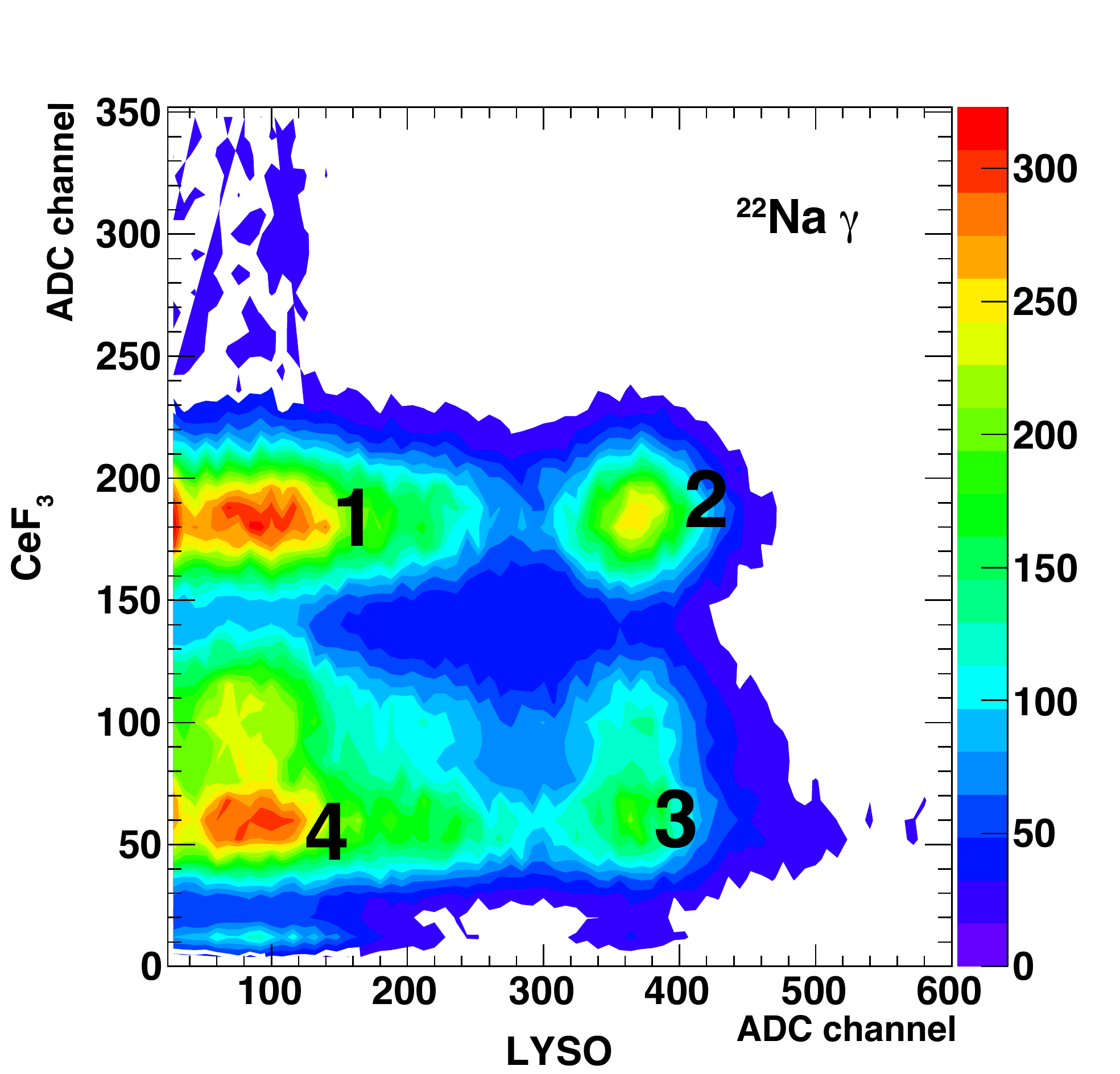}\hspace{2pc}%
\end{figure}

Direct scintillation signals from $\gamma$ quanta depositing energy in the LYSO bar itself have also been studied, since the LYSO bar volume is $\sim25$\% of  the CeF$_3$ crystal one. Indeed, they have been easily visualised. For this purpose, the signals from the PMT coupled to the LYSO bar have been attenuated by 12 db. The correlation is found in Fig.~\ref{f-LYSOscint}, where the numbers indicate the different topologies that are observed for 511 keV $\gamma$: (1) photoelectric effect in CeF$_3$ and Compton background in LYSO, (2) photoelectric effect in both, CeF$_3$ and LYSO, (3) Compton background in CeF$_3$ and photoelectric effect in LYSO and (4) Compton background in both, CeF$_3$ and LYSO. 

Overall, the studies described in this section, making use of a relatively big LYSO bar coupled to a CeF$_3$ crystal, have allowed to commission the setup and to learn about the performance of the basic elements involved. Photoluminescence from LYSO, excited by the CeF$_3$ scintillation emission, has been observed, with cosmic $\mu$ and $^{22}$Na $\gamma$ quanta leading to a consistent WLS-to-direct signal amplitude ratio. Furthermore, as expected, a direct scintillation signal from $^{22}$Na $\gamma$ quanta interacting in the LYSO bar was observed. The lesson to be learned from these tests is, that for all candidate WLS materials and their volumina, the possible presence of direct scintillation needs to be always studied and its tolerable amplitude optimised in the design.

\section{Cerium-doped quartz fibres as WLS candidates}
\label{s-CeSiO2}
\begin{figure}[!t]
\includegraphics[clip=true, trim= 0cm 8mm 5mm 0cm,width=14pc]{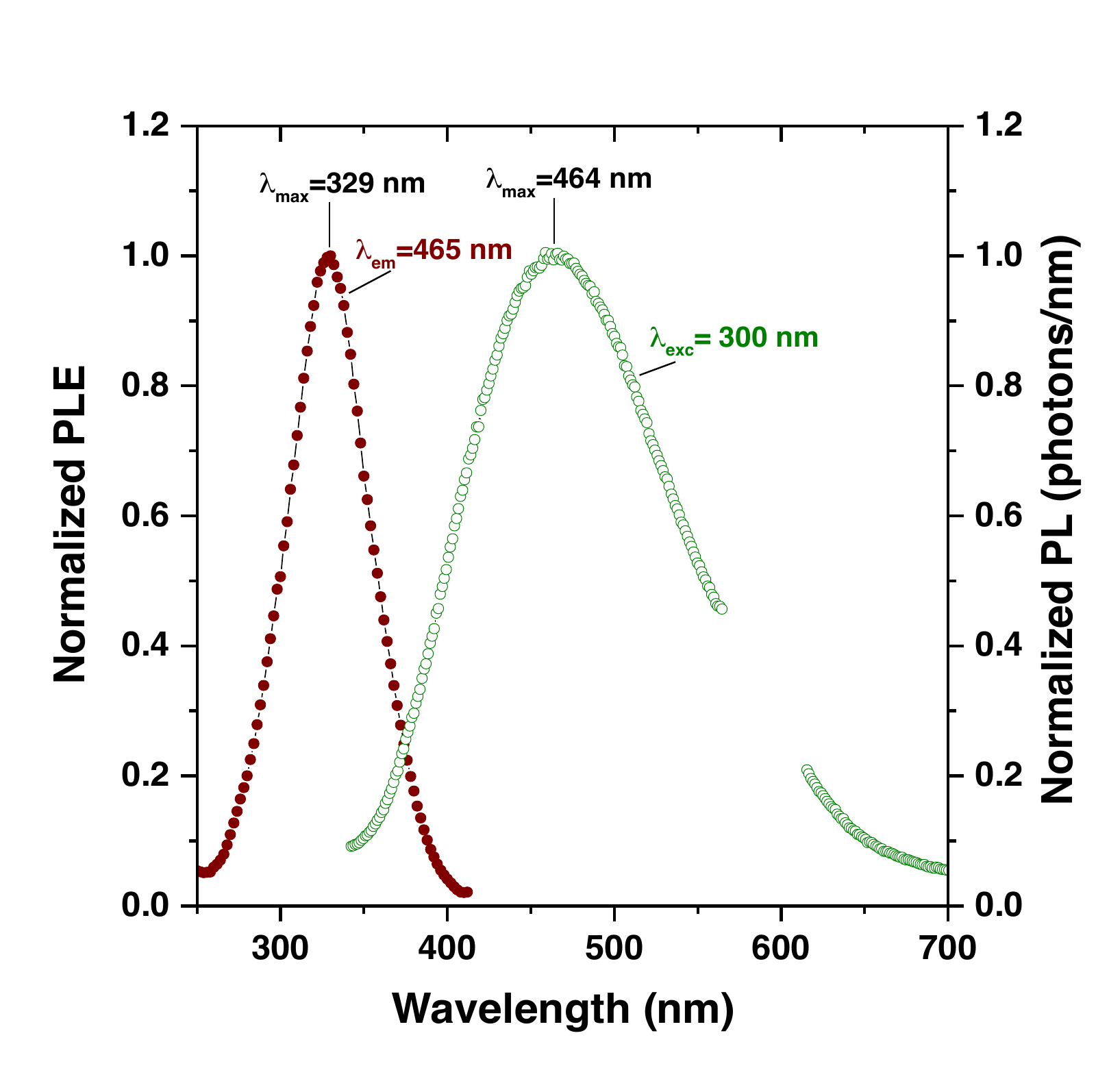}\hspace{2pc}%
\begin{minipage}[b]{22pc}\caption{\label{f-QuartzEmission} Photoluminescence emission and excitation curves for cerium-doped quartz fibres used in this test~\cite{r-vedda,r-PLVED}.}
\end{minipage}
\end{figure}
\begin{figure}[!b]
\begin{center}
 \begin{tabular}[h]{cc}
{\mbox{\includegraphics[width=70mm]{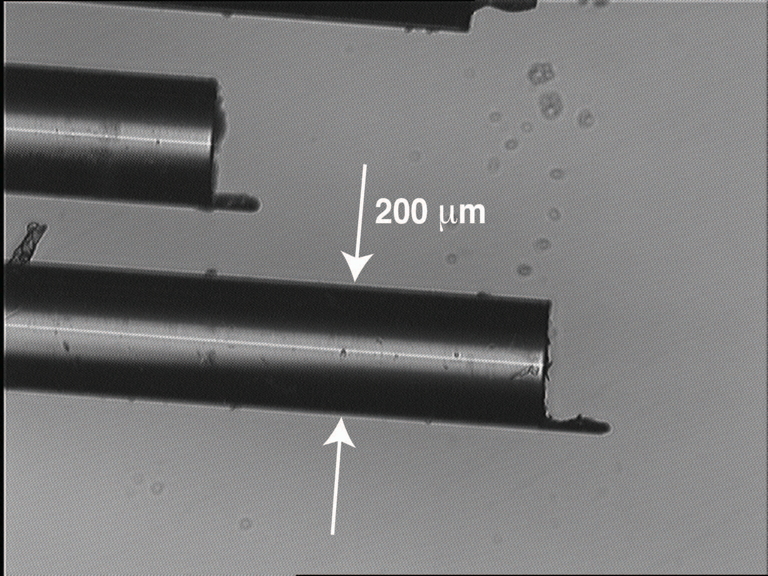}}} &
{\mbox{\includegraphics[width=57mm]{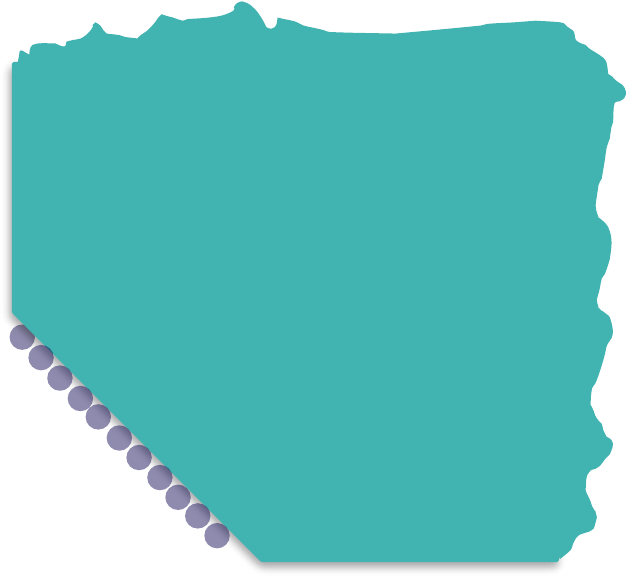}}}
   \end{tabular}
\end{center}
\caption{Image of the Ce-doped quartz fibre ends, as seen under the microscope (left) and concept view of the fibre arrangement along the 3 mm wide crystal chamfer (right).}
\label{f-fibres}
\end{figure}

As a material in calorimetry and dosimetry, cerium-doped quartz is known to be quite radiation-hard, a feature that makes it an interesting WLS candidate for HL-LHC applications. For this reason, a first attempt at extracting a WLS signal from such fibres has been performed.
The photoluminescence of the fibres considered in this test matches the emission spectrum of CeF$_3$~\cite{r-PLVED}, as visible in Fig.~\ref{f-QuartzEmission}. A fast emission component, as needed for HL-LHC applications, had been previously observed~\cite{r-vedda}. 
A set of fibres was received, that were developed for dosimetry applications, and as such had been optimised for radioluminescence~\cite{r-VEDCHIO}.
These fibres had not been optimised yet for photoluminescence, nor had its fast emission component been maximised. Furthermore, they had been grown in an oxidising atmosphere, and were thus rich in Ce$^{4+}$, which is optically inactive but highly absorbing, while poor in photoluminescent Ce$^{3+}$ content. The fibres received had a diameter of 200 $\mu$m, they had no cladding, its end were, out of necessity, simply cut with a razor blade and were not polished neither aluminised (Fig.~\ref{f-fibres}). To maximise the light collection, 12 fibres, 10 cm long, were laid along the crystal chamfer in a single-layer ribbon (Fig.~\ref{f-fibres}) replacing the LYSO bar in setup 2 (Fig.~\ref{f-setup}). An air gap of a few millimetres was left between the ends of the fibres and the PMT window. A modest, if at all, signal was expected. 
\begin{figure}[!t]
\begin{center}
 \begin{tabular}[h]{ccc}
{\mbox{\includegraphics[clip=true, trim= 0cm 87mm 0cm 0cm,width=50mm]{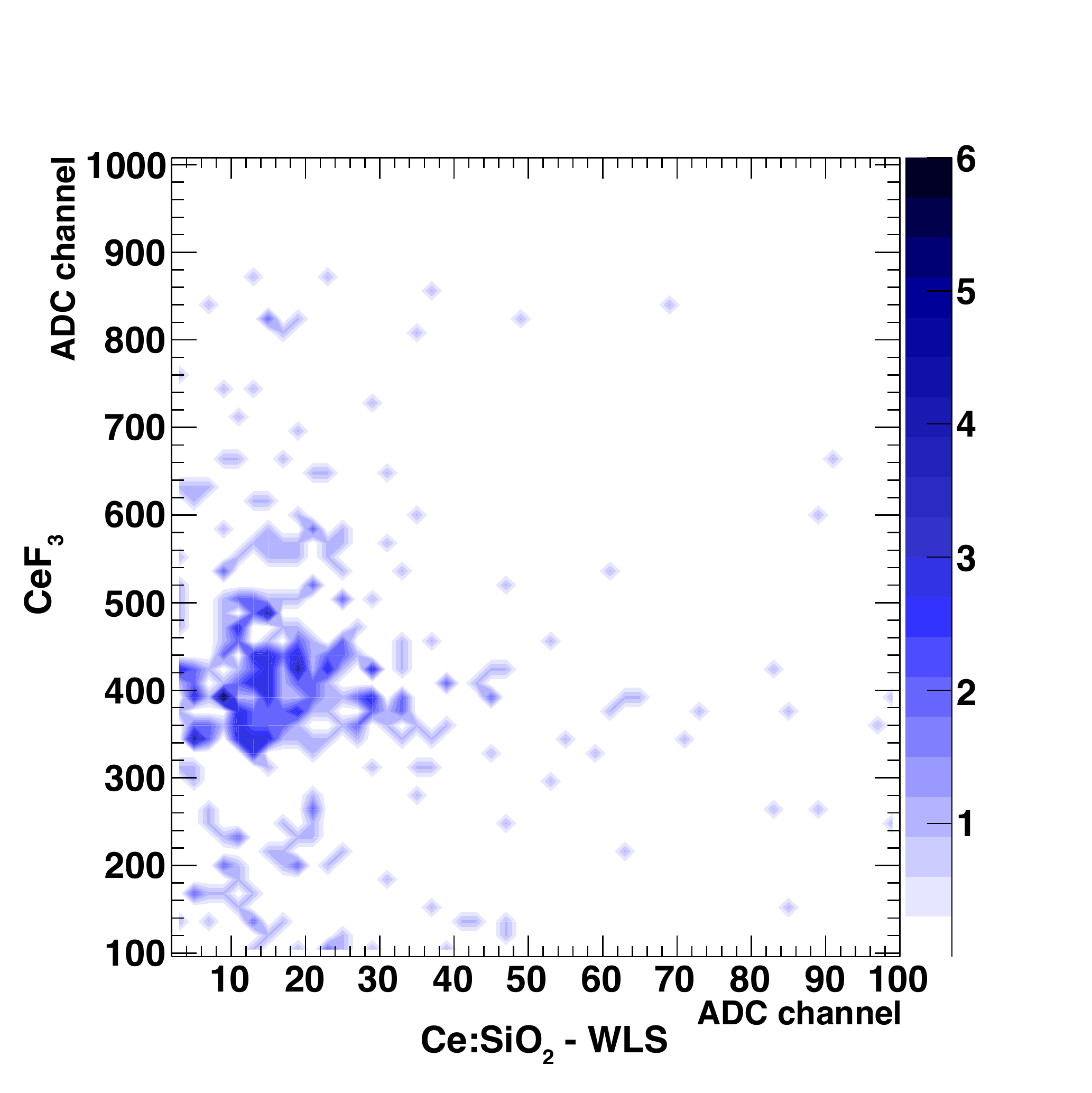}}} &
{\mbox{\includegraphics[width=50
mm]{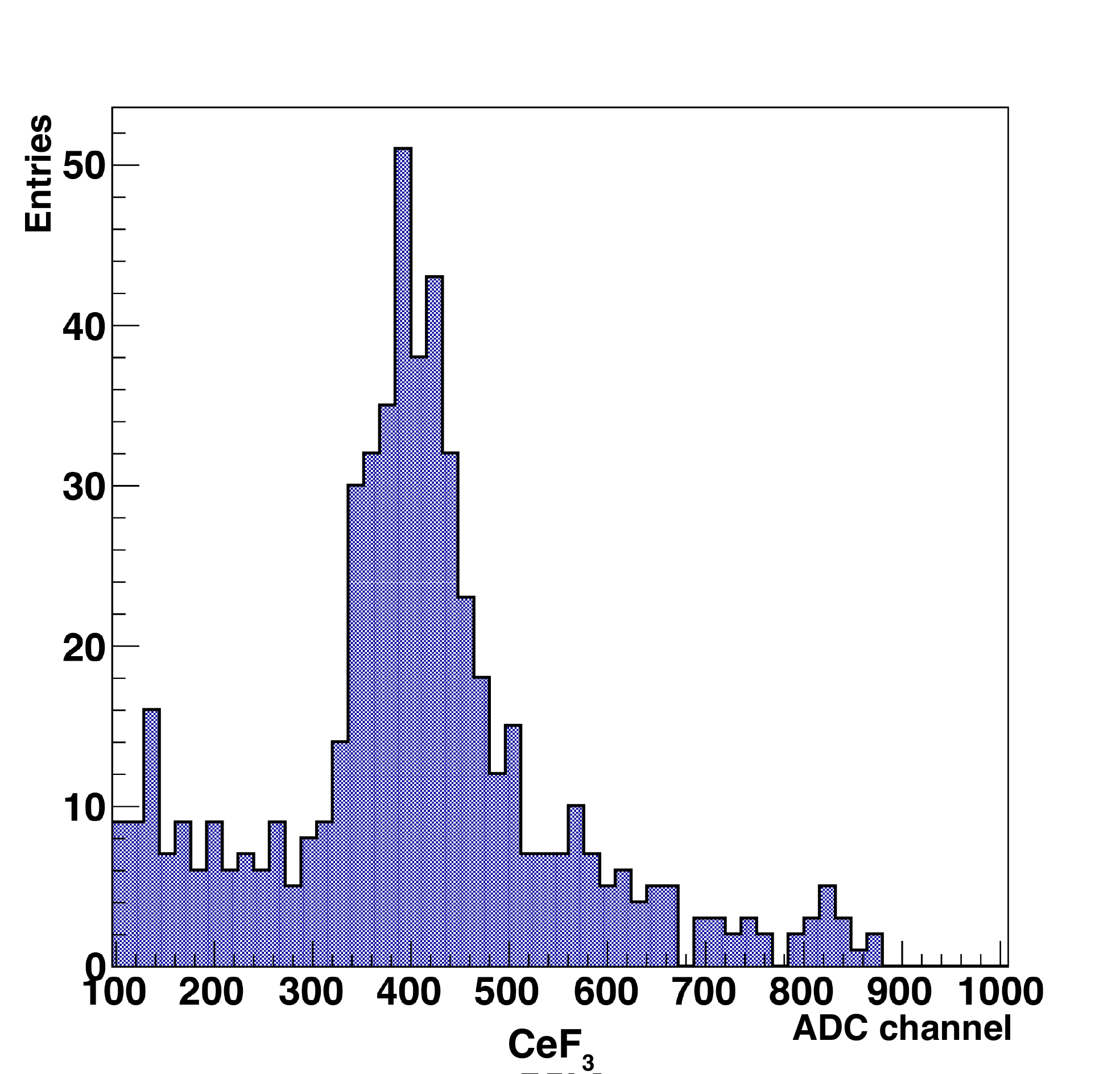}}} &
{\mbox{\includegraphics[width=50mm]{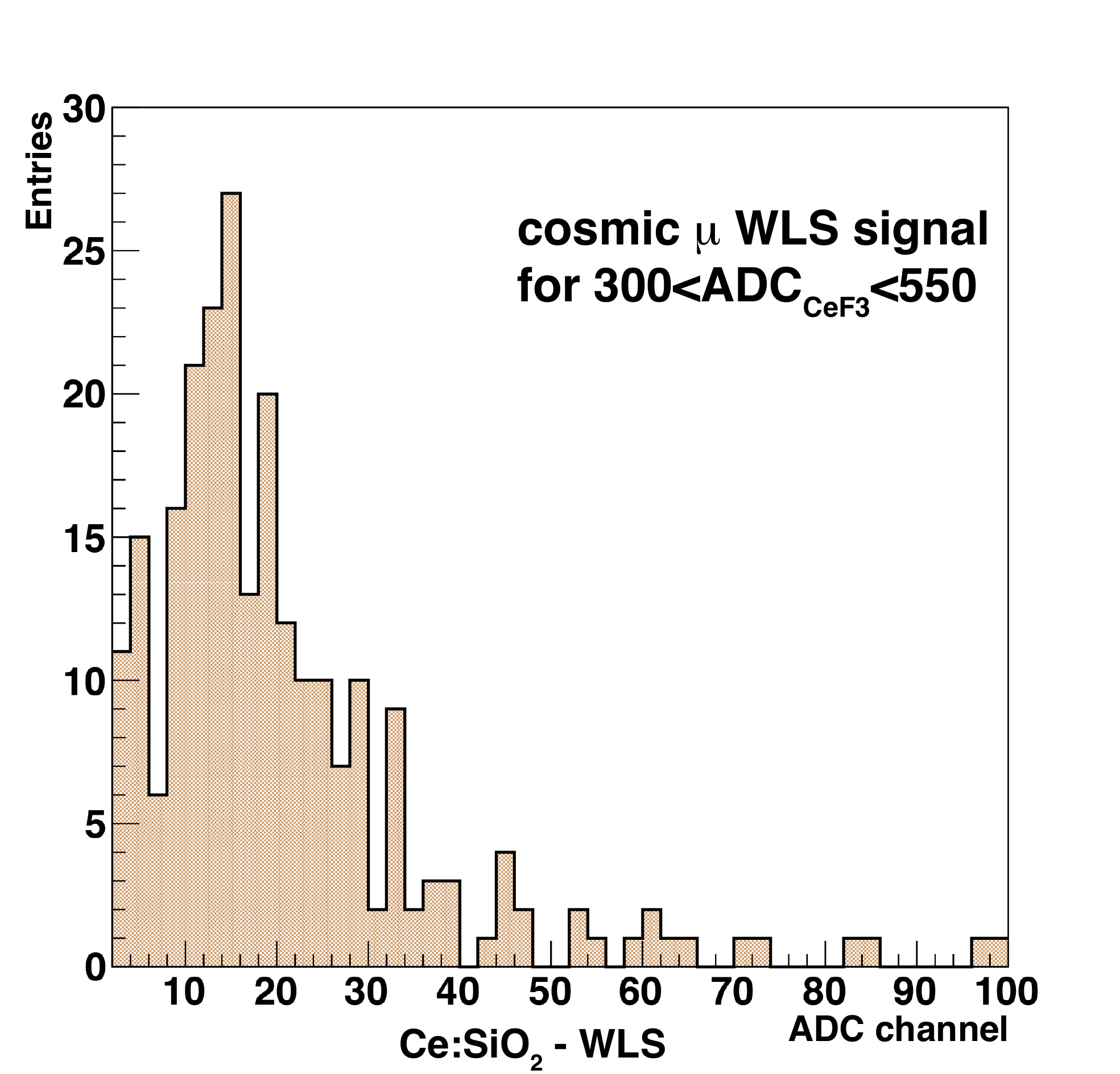}}}
   \end{tabular}
\end{center}
\caption{For cosmic muons  in setup 2 (Fig.~\ref{f-setup}), correlation between direct scintillation signal from the bottom CeF$_3$ and WLS signal from a ribbon of Ce-doped quartz fibres coupled to the top CeF$_3$ crystal, and one-dimensional projections where, for the WLS signal spectrum, amplitudes corresponding to $300<{\mathrm{ADC_{CeF_3}}}<550$ were required.}
\label{f-quartz1crystal}
\end{figure}
The observed correlation for cosmic muons, between direct scintillation signal from CeF$_3$, and WLS signal from the ribbon of fibres, is visible in Fig.~\ref{f-quartz1crystal}, together with one-dimensional projections. For the WLS signal, the direct scintillation amplitude was required to fall between ADC channels 300 and 550. The ratio of direct--to--WLS light detected was 0.3\% for this setup. Although the ratio is very modest, taking into account all the limitations of the present test, it can be considered promising in view of an optimisation of such fibres characteristics. 
Due to the total fibre volume of only $0.037$ cm$^3$,  the direct scintillation signal in the fibres is very small, and remained undetectable in this setup.  

\begin{figure}[!h]
\begin{center}
 \begin{tabular}[h]{ccc}
{\mbox{\includegraphics[clip=true, trim= 7mm 0mm 10mm 0cm,width=45mm]{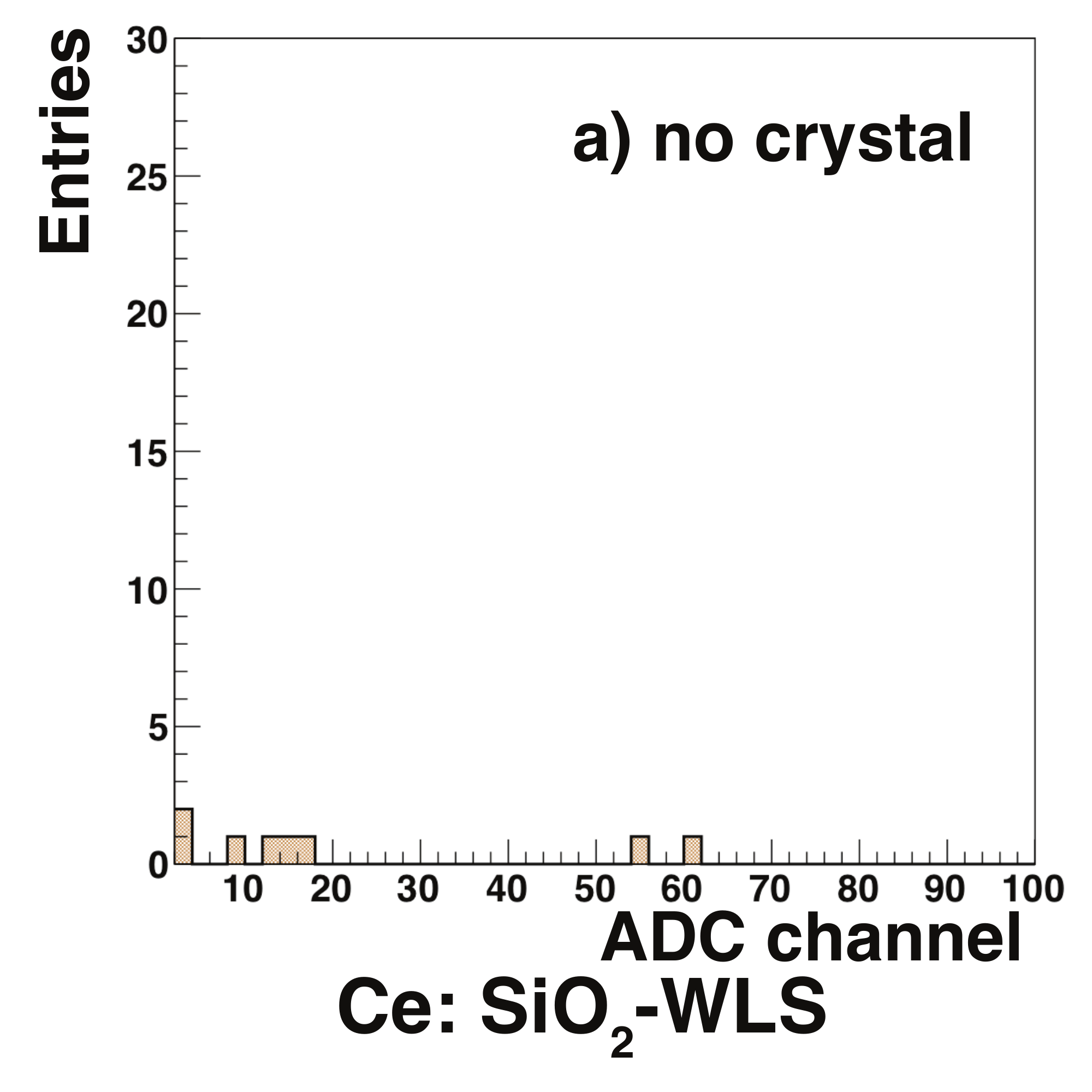}}} &
{\mbox{\includegraphics[clip=true, trim= 7mm 0mm 10mm 0cm,width=45mm]{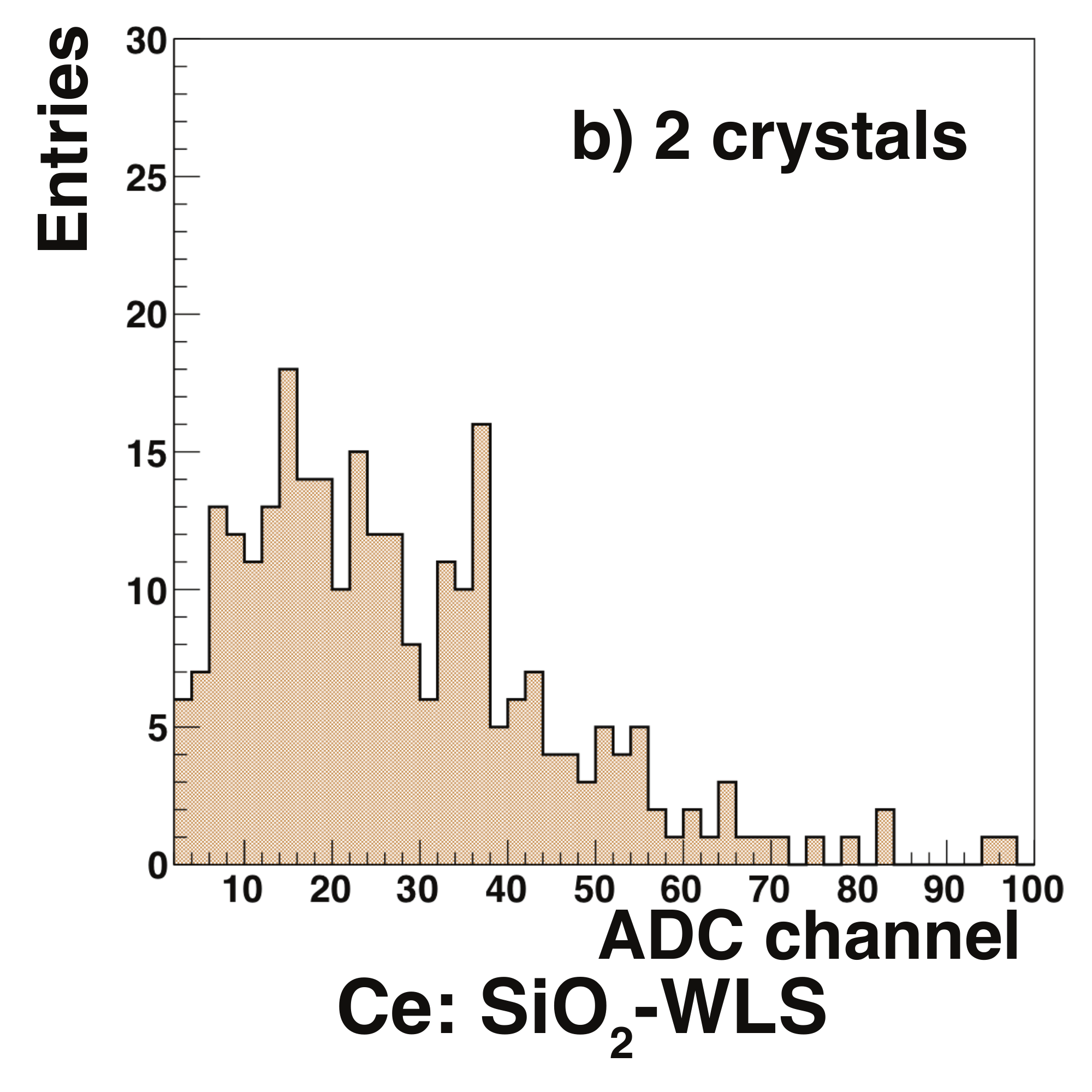}}} &
{\mbox{\includegraphics[clip=true, trim= 7mm 0mm 10mm 0cm,width=45mm]{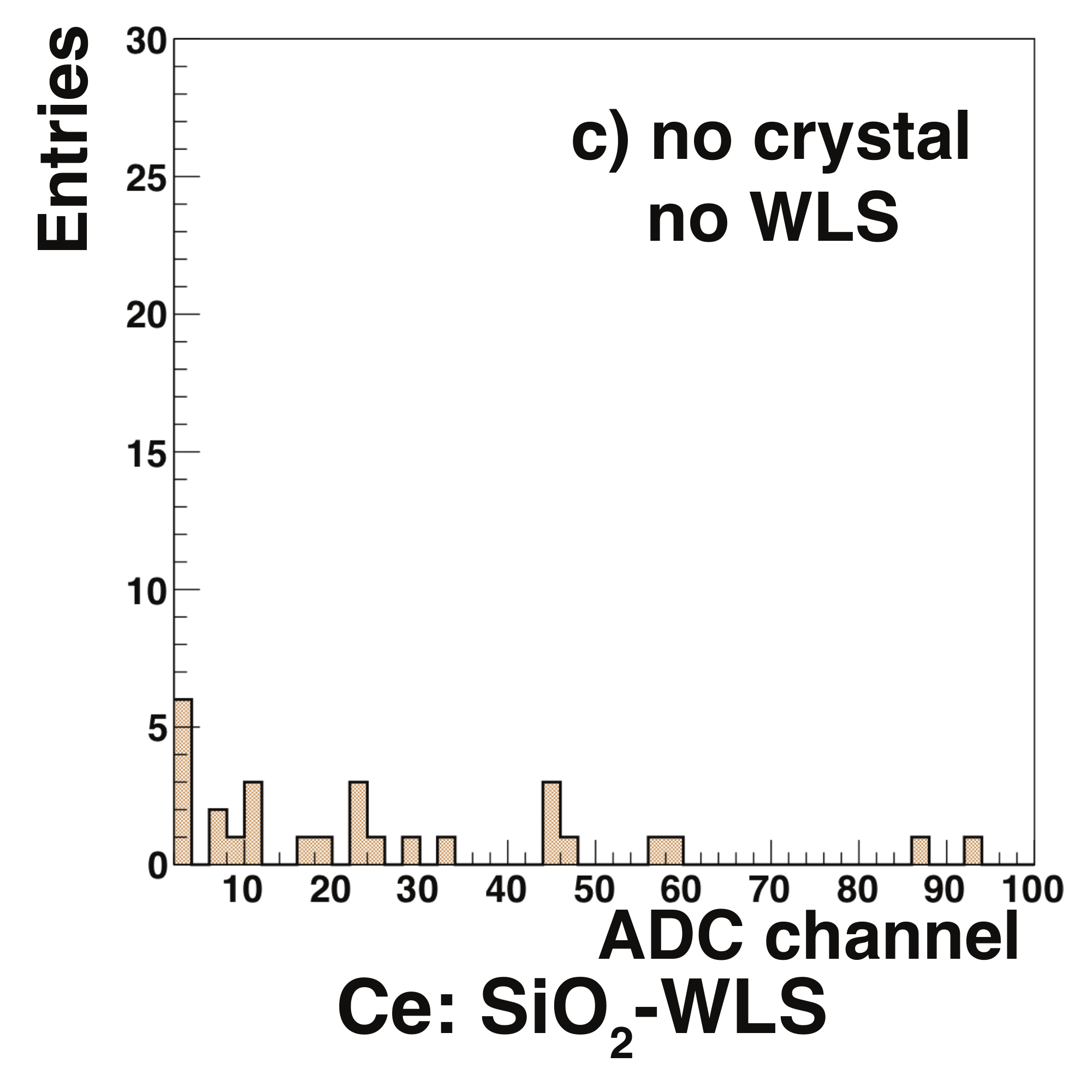}}} \\
\end{tabular}
\begin{tabular}[h]{cc}
{\mbox{\includegraphics[clip=true, trim= 7mm 0mm 10mm 0cm,width=45mm]{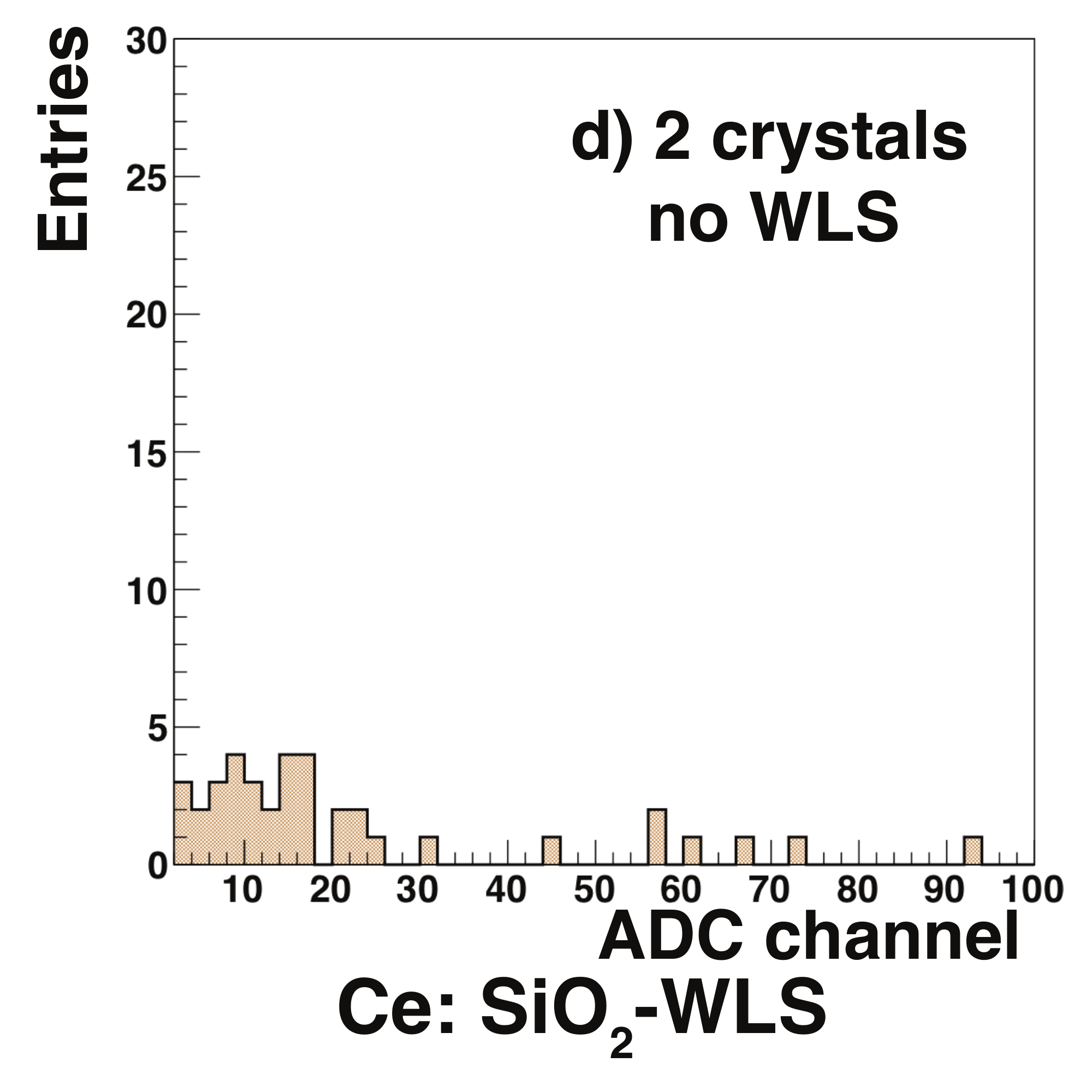}}} &
{\mbox{\includegraphics[clip=true, trim= 7mm 0mm 10mm 0cm,width=45mm]{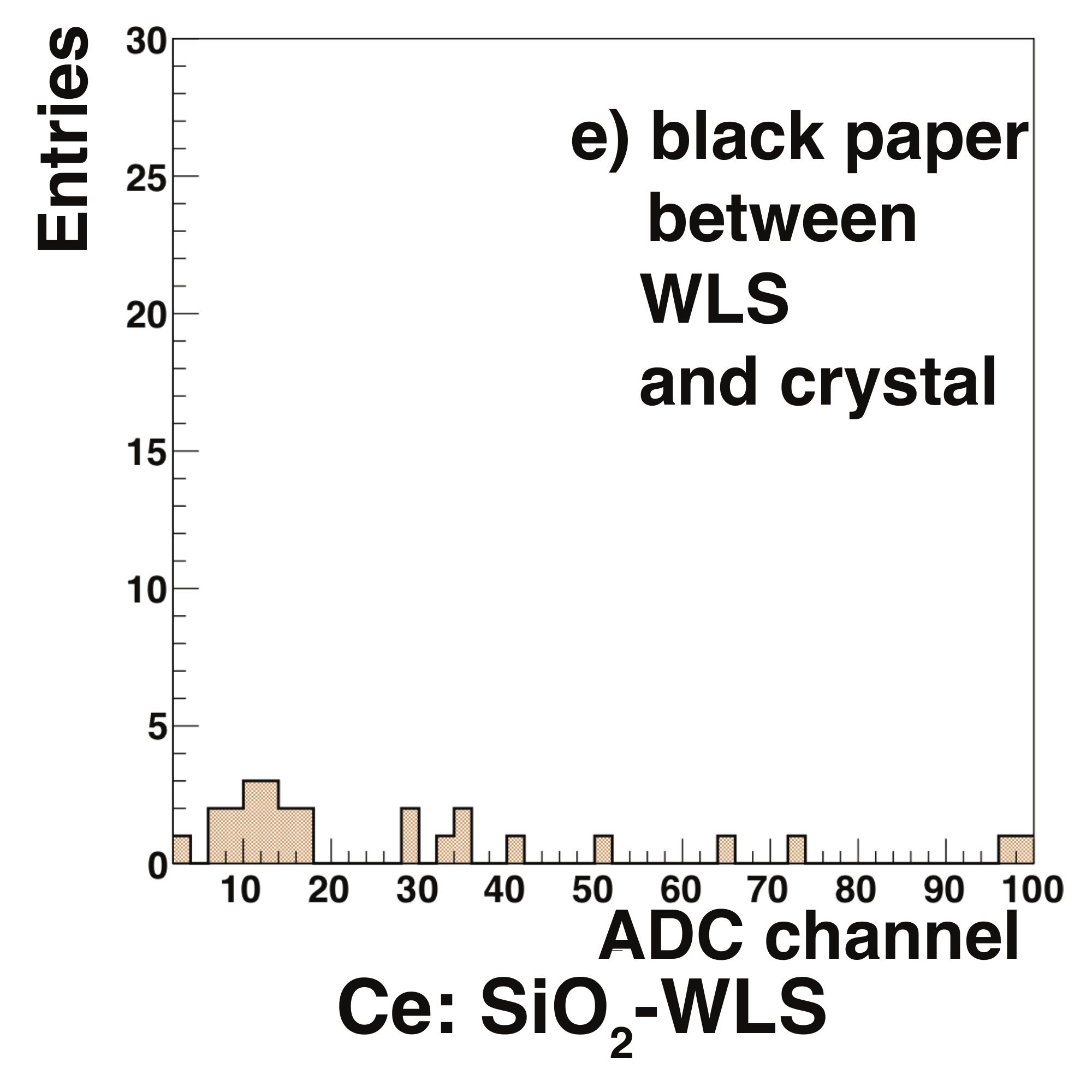}}} \\
\end{tabular}
\end{center}
\caption{For the same duration of data taking for cosmic muons, WLS signal spectra for Ce-doped quartz fibres in setup 2 (Fig.~\ref{f-setup}), for several cross-check configurations to the results in Fig.~\ref{f-quartz1crystal}: a) crystal removed; b) 2 crystals coupled to the fibres ribbon; c) no crystal, no fibres present; d) 2 crystals, but no fibres present and e) crystal optically isolated from the WLS fibres.}
\label{f-crosschecks}
\end{figure}

A number of cross-check measurements were performed, always for the same duration of data taking, to ascertain that the observed correlation is indeed due to a WLS signal. The results match the expectations.
In the test whose resulting WLS signal spectrum is visible in Fig.~\ref{f-crosschecks}a), the crystal coupled to the WLS fibres ribbon was removed from the setup: the spectrum is compatible with no WSL signals detected.
In Fig.~\ref{f-crosschecks}b), the WLS signal spectrum was taken for two crystals, sitting on top of each other, both coupled to the WLS fibres ribbon. The mean of the distribution is higher and its width broader than for a single crystal, although slightly different from the exact convolution of two identical distributions as of Fig.~\ref{f-quartz1crystal}, due to differences in chamfer dimensions and crystal light yield. When the crystals and also the WLS fibres ribbon were removed (Fig.~\ref{f-crosschecks}c), the distribution was found to be compatible with no signal. Similarly, this is also the case when only the WLS fibres were removed (Fig.~\ref{f-crosschecks}d) and when a black paper stripe was inserted between the crystal chamfer and the WLS fibres ribbon (Fig.~\ref{f-crosschecks}e). The results of these cross-checks measurements confirm the observation of Fig.~\ref{f-quartz1crystal}, that a WLS signal is observed, for a cerium fluoride crystal coupled to Ce-doped quartz fibres.
\begin{figure}[!t]
\includegraphics[clip=true, trim= 0mm 12mm 0mm 0cm,width=12pc]{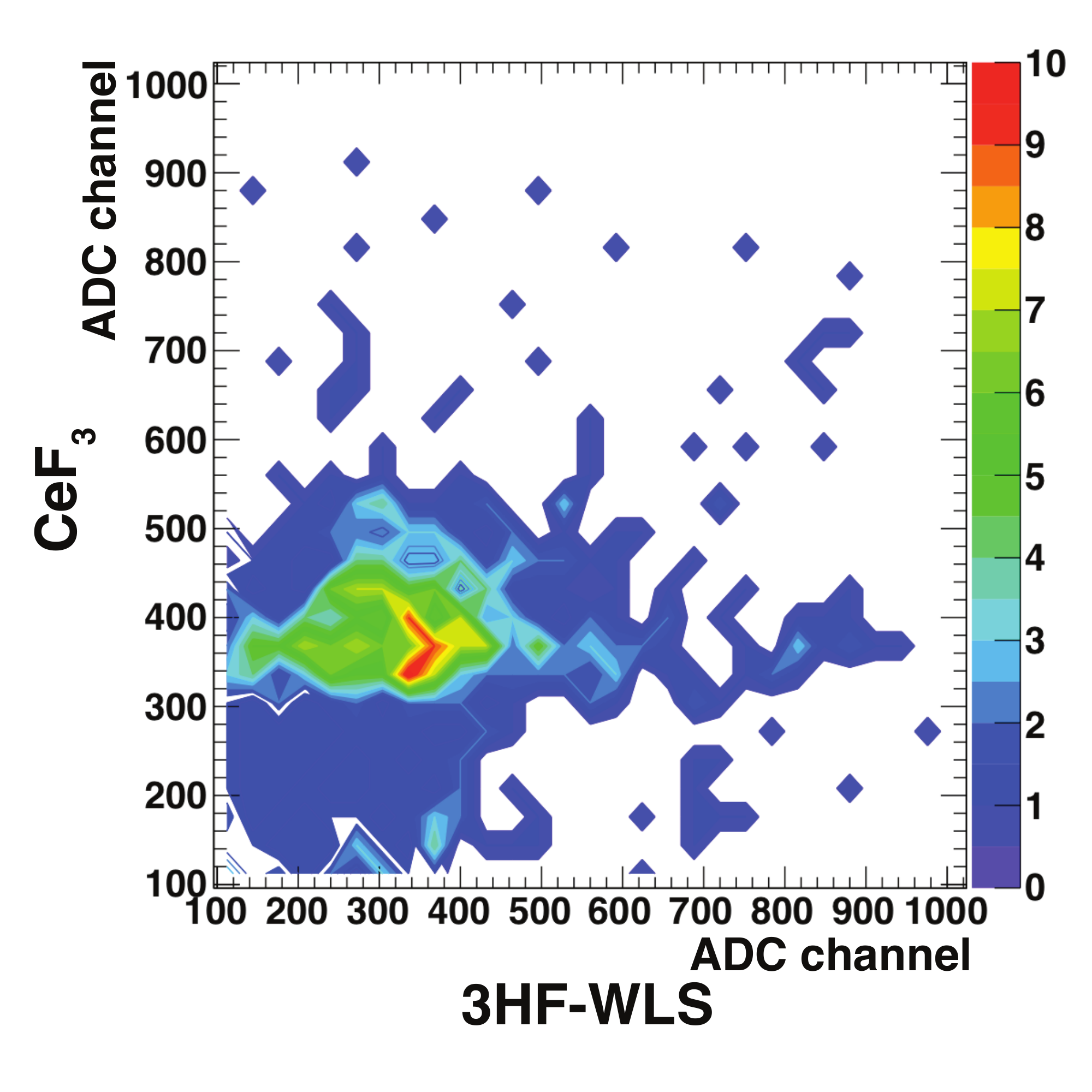}\hspace{2pc}%
\begin{minipage}[b]{24pc}\caption{\label{f-3HF}For cosmic muons in setup 2 (Fig.~\ref{f-setup}), correlation between direct scintillation signal from CeF$_3$ and WLS signal from a Kuraray 3HF fibre.}
\end{minipage}
\end{figure}

\section{Tests of 3HF Kuraray fibres as WLS}
\label{s-kuraray}
The construction of a prototype calorimeter cell, to be tested in a particle beam, following the proposed readout concept, was being prepared.
To allow for a WLS signal readout, while R\&D is being performed on Ce-doped quartz fibres, the temporary use of conventional plastic WLS fibres was envisaged.
A suitable candidate was found in 3HF fibres from Kuraray, that are known to be photoluminescent  over the range of cerium fluoride scintillation emission wavelengths~\cite{r-KURARAY,r-KURPRIV}. Single-clad fibres, 1 mm in diameter, were chosen, since the cladding material, consisting of PMMA, is known to be at least partially transmitting over the wavelength range of the cerium fluoride scintillation emission~\cite{r-MIL}, while no data were available for the proprietary outer cladding used on double-clad fibres, which consists of a fluorinated polymer.
For this first test, a single 3HF fibre, 1 mm in diameter, single-clad, with no aluminised end, was used. An air gap was present between the fibre and the PMT.
The correlation in Fig.~\ref{f-3HF} shows the detection of a WLS signal. The ratio of direct-to-WLS light output is at the level of 0.5\%, an amplitude which corresponds to expectations, due to the measurement conditions, which have not been fully optimised yet.

\section{Prototype cell construction and commissioning}
\label{s-proto}
\begin{figure}[!b]
\centering
\includegraphics[clip=true, trim= 5mm 7mm 2mm 0cm,width=12cm]{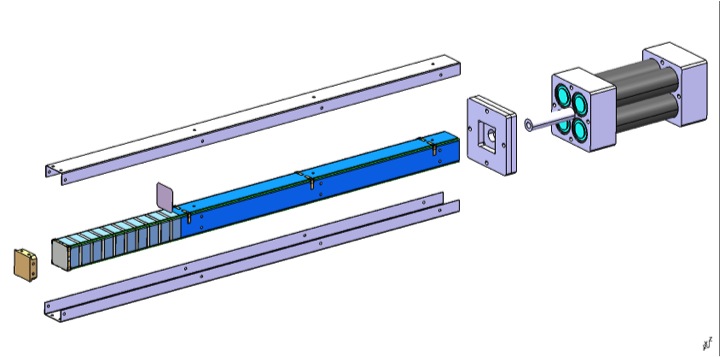}
\caption{Concept drawing for a complete calorimeter sampling cell, in an exploded view.}
\label{f-explodedchannel}
\end{figure}
\begin{figure}[!t]
\begin{center}
 \begin{tabular}[h]{ccc}
{\mbox{\includegraphics[width=37mm]{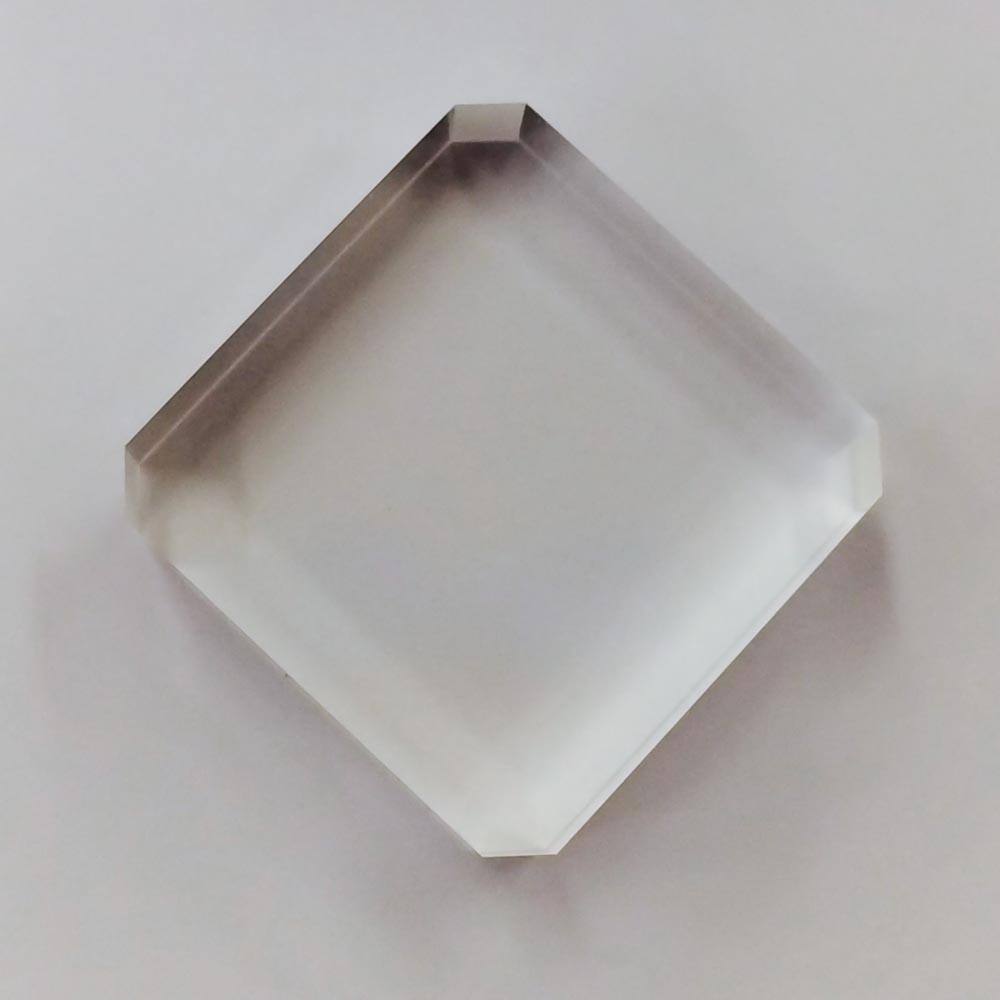}}} &
{\mbox{\includegraphics[width=50mm]{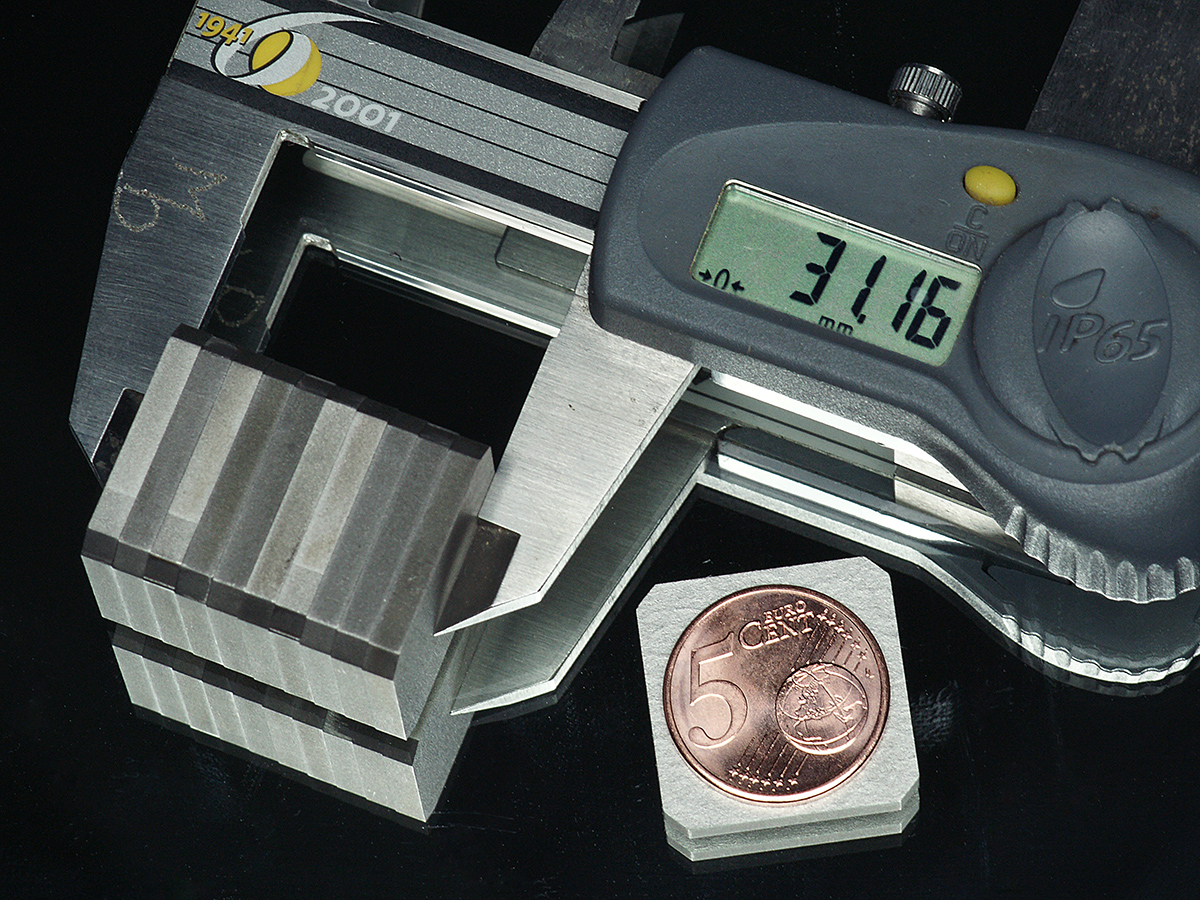}}} &
{\mbox{\includegraphics[width=50mm]{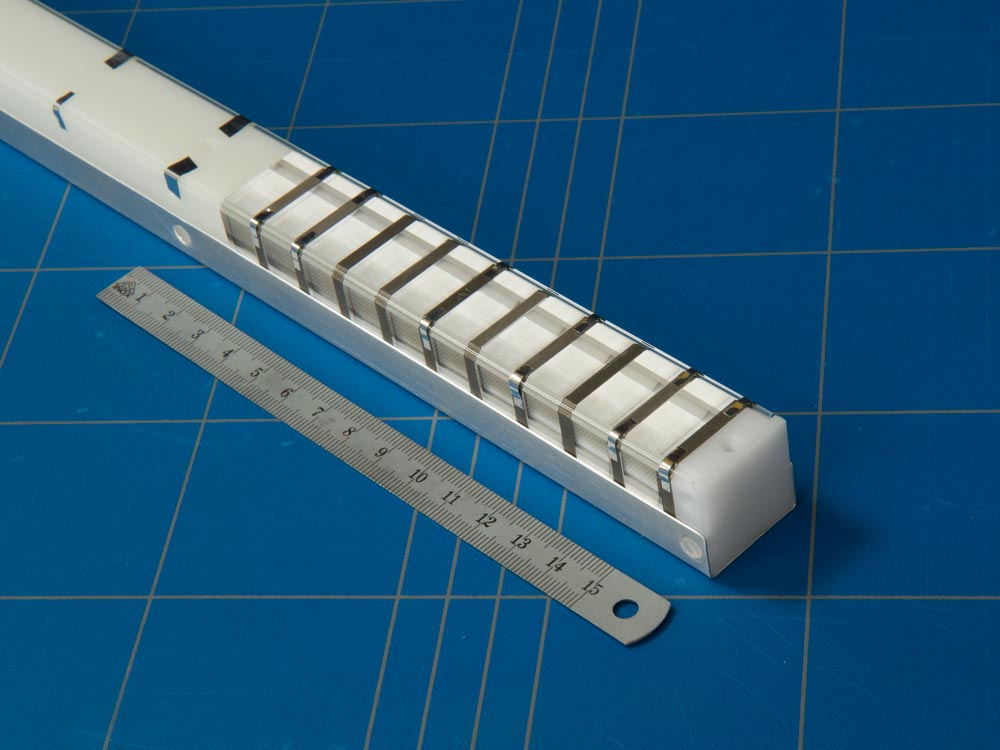}}}
   \end{tabular}
\end{center}
\caption{Prototype calorimeter cell construction elements: single cerium fluoride crystal (left), set of tungsten plates (centre) and prototype during assembly, showing the ten samplings and the WLS fibres running along the edges (right).}
\label{f-parts}
\end{figure}
Based on the successful proof-of-principle measurements described above, a prototype cell has been constructed for tests in high-energy particle beams,
whose 3-dimensional drawing is presented in Fig.~\ref{f-explodedchannel}.
The dimensions were optimised for a $\sim500$ MeV electron beam. Ten samplings were foreseen, with tungsten absorber plates of 3.1 mm thickness and cerium fluoride crystals of 10 mm thickness (Fig.~\ref{f-parts}), for a total of 17 $\mathrm{X}_0$, while transverse dimensions of $24\times 24\;{\mathrm{mm}}^2$ were chosen. 
It should be noticed that for this configuration the effective Moli\`ere radius is $R_M=23$~mm, but a smaller value can be easily obtained through a different choice of relative absorber and scintillator thickness. For example, for $2.5$ mm thick tungsten plates and $1.5$ mm thick cerium fluoride crystals, $R_M=14$ mm is achievable.

The mechanical support structure was conceived to be easily expandable in depth for beam tests at higher energies. The samples were all provided with 3 mm chamfers along the four 10 mm long corners, and a Kuraray 3HF single-clad fibre was running along each one of the long cell edges. For the test, each fibre was coupled to a Hamamatsu PMT with a bialkali photocathode of type R1450, within a H6524 assembly, and it was aluminised at the end opposite to the photodetector. 

\begin{figure}[!h]
\includegraphics[clip=true, trim= 0mm 3mm 0mm 0cm,width=12pc]{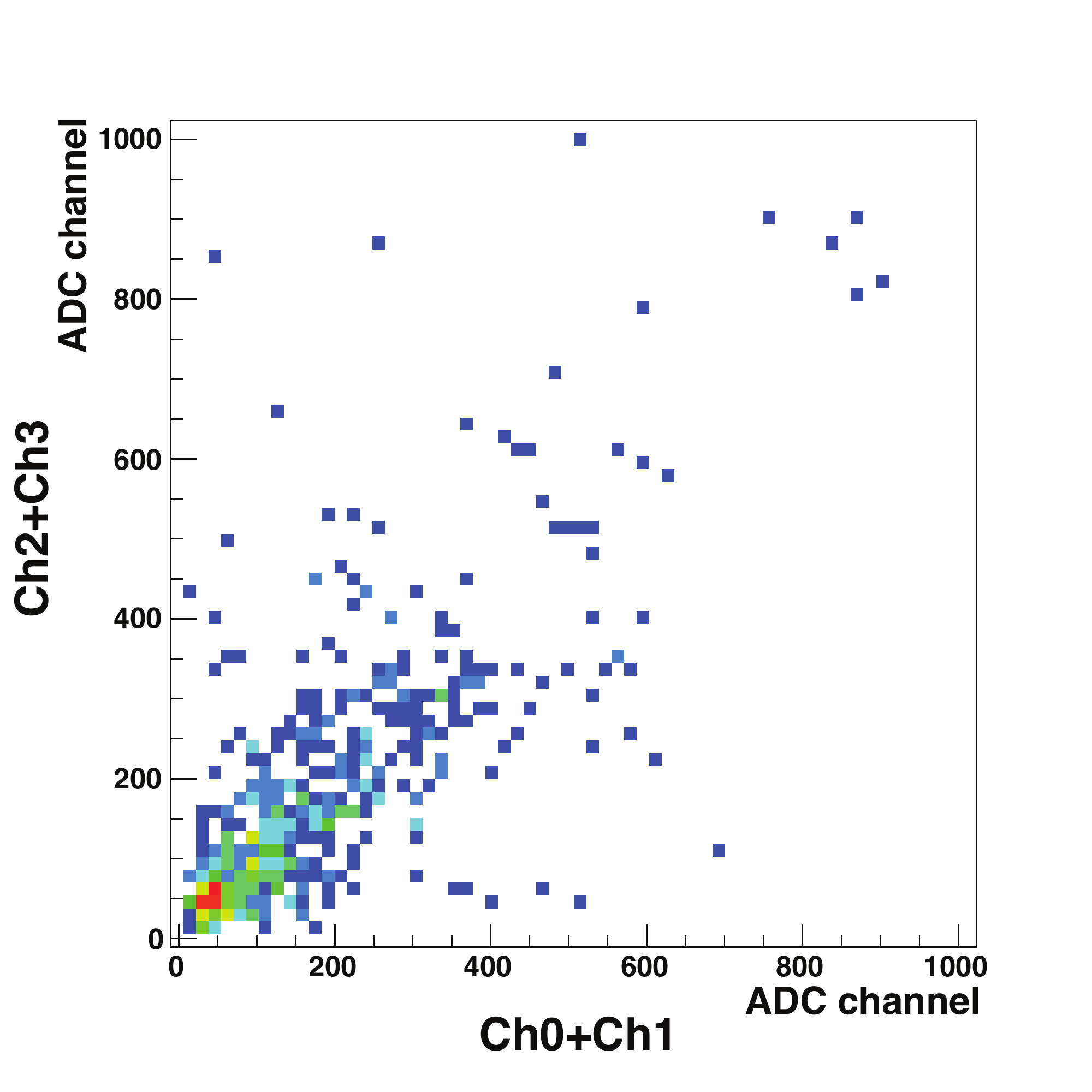}\hspace{2pc}%
\begin{minipage}[b]{24pc}\caption{\label{f-cosmicschannel}For cosmic muons going through the length  of the prototype cell of Fig.~\ref{f-explodedchannel}, correlation between the two sums of WLS signals from fibres running along two diagonally opposite edges.}
\end{minipage}
\end{figure}

Prior to an exposure to high-energy particle beams, the prototype was commissioned in the laboratory, by acquiring signals from cosmic muons required to traverse the cell along its length. This was achieved by placing the cell vertically upright, and triggering with scintillators placed on top and below the cell. Muons going only partially through the length of the cell were not excluded, due to a trigger counter size slightly larger than the transverse cell dimensions.
The correlation between sums of signals coming from WLS fibres running along diagonally opposite edges is found in Fig.~\ref{f-cosmicschannel}. This being raw data, the slight deviation from a unitary slope is due to the differences in quantum efficiency folded with the gain of the photomultiplier tubes. The measurement allowed to ascertain the functioning of all components and to make sure, that signals of detectable amplitude are generated.

\section{Conclusions}
\label{s-conc}
This paper summarises the work that has been performed so far, in pursuing cerium fluoride scintillating crystals used in a simplified sampling calorimeter geometry. The proposed concept has wavelength-shifting fibres running along the four edges of the stack, collecting light through slightly depolished crystal chamfers.

In proof-of-principle laboratory tests, wavelength-shifted signals have been detected for conventional fibres from Kuraray, and for Ce-doped quartz fibres.
While the fibres need to be further optimised  for HL-LHC calorimetry applications, a full prototype cell has been built and commissioned in the laboratory with cosmic muons, before tests in a high-energy beam.

The presented measurements demonstrate, that this principle with its components has the potential to be used for a high-rate, radiation resistant sampling calorimeter, as needed for HL-LHC upgrades, in a mechanical assembly that allows a simple construction.

\section*{Acknowledgments}
The excellent support of the ETH Zurich technical staff is acknowledged, in particular M. Dr\"oge, C. Haller and U. Horisberger, as is the one of CERN, by R. Dumps, T. Schneider and M. Van Stenis. We are grateful to A. Vedda, N. Chiodini and M. Fasoli (U. Milano Bicocca), and to Kuraray (Japan) for kindly providing us with samples for this test. We also acknowledge the effort made by Hamamatsu (Japan) and Tokuyama (Japan), to provide us with the needed parts in time for meeting a stringent construction schedule. This work was performed with the support of the Swiss National Science Foundation. 
\\





\begin{thebibliography}{1}
\bibitem{r-LUC} M.~Lucchini (CMS Collab.), ``Evolution of the response of the CMS ECAL'', Proc. IEEE Nuclear Science Symposium (Seoul, S. Korea, 2013) and CMS-CR-2013-401 (CERN, Geneva, Switzerland).
\bibitem{r-NIMCEF3} G.~Dissertori, P.~Lecomte, D.~Luckey, F.~Nessi-Tedaldi, F.~Pauss, 
  Th.~Otto,  S.~Roesler, Ch.~Urscheler,
Nucl.~Instr.~and Meth.~Phys.~Res.~A 622 (2010) 41-48.
\bibitem{r-ZHULYSO} J.~Chen, R.~Mao, L.~Zhang, R.-Y.~Zhu, IEEE Trans.~Nucl.~Sci.~54, (2007) 1319-1326.
\bibitem{r-vedda} A.~Vedda, N.~Chiodini, D.~Di Martino, M.~Fasoli, S.~Keffer, A.~Lauria, M.~Martini, F.~Moretti, G.~Spinolo, M.~Nikl, N.~Solovieva and G.~Brambilla, Appl.~Phys.~Lett., Vol.~85 (2004) 6356.
 \bibitem{r-AND2} D.~F.~Anderson, Nucl.~Instr.~Meth A 287 (1990) 606-612.
\bibitem{r-EACEF3} E.~Auffray, S.~Baccaro, T.~Beckers, Y.~Benhammou, A.~N.~Belsky, B.~Borgia, D.~Boutet, R.~Chipaux, et al., Nucl.~Instrum.~and Meth.~Phys.~Res.~A 383 (1996) 367-390.
\bibitem{r-LTNIM} M.~Huhtinen, P.~Lecomte, D.~Luckey, F.~Nessi-Tedaldi, F.~Pauss, Nucl.~Instr.~and Meth.~A 545 (2005) 63-87.
\bibitem{r-NIMLYSO} G.~Dissertori, D.~Luckey, F.~Nessi-Tedaldi, F.~Pauss, M.~Quittnat, R.~Wallny, Nucl.~Instr.~and Meth.~Phys.~Res.~A 745 (2014), 1-6.
\bibitem{r-PLVED} A.~Vedda, N.~Chiodini, M.~Fasoli, private communication.
\bibitem{r-VEDCHIO} Courtesy of A.~Vedda and N.~Chiodini, University of Milano-Bicocca (Italy).
\bibitem{r-VEDRED} M.~Fasoli, A.~Vedda, A.~Lauria, F.~Moretti, E.~Rizzelli, N.~Chiodini, F.~Meinardi, M.~Nikl, J.~Non-Cryst.~Sol.~355 (2009) 1140-1144.
\bibitem{r-KURARAY} Kuraray, Japan, 3HF scintillation fibres technical specifications at http://kuraraypsf.jp/psf/sf.html, visited on May 15th, 2014.
\bibitem{r-KURPRIV} Kuraray, private communication.
\bibitem{r-MIL} D.~C.~Miller et al., Proc.~SPIE Solar Energy and Technology Conference, San Diego (USA, 2009).

\end{thebibliography}
%

\end{document}